\documentclass[pre,float,twocolumn,showpacs,superscriptaddress]{revtex4}

\usepackage{float}
\usepackage{mathtools}
\usepackage{amsfonts}
\usepackage{amsmath}
\usepackage{graphicx}
\usepackage[usenames,dvipsnames]{color}
\usepackage{amssymb}
\usepackage{longtable}
\usepackage{cancel}
\usepackage{undertilde}
\usepackage{bm} 
\usepackage{bbm} 
\usepackage{tabularx}
\usepackage{makeidx,xspace,color,listings}

\allowdisplaybreaks

\newcommand{\beq}{\begin{equation}} 
\newcommand{\eeq}{\end{equation}}
\newcommand{\bqa}{\begin{eqnarray}} 
\newcommand{\eqa}{\end{eqnarray}}
\newcommand{\nn}{\nonumber}

\newcommand{\dg}{^\dagger}
\newcommand{\rt}[1]{\sqrt{#1}\,}

\newcommand{\bra}[1]{\langle{#1}|} 
\newcommand{\ket}[1]{|{#1}\rangle}

\newcommand{\op}[2]{|{#1}\rangle \langle{#2}|}

\newcommand{\RPone}{\text{${\rm RP_1}$}}
\newcommand{\RPtwo}{\text{${\rm RP_2}$}}
\newcommand{\RPthr}{\text{${\rm RP_3}$}}

\newcommand{\mapK}{{\cal K}}

\newcommand{\Hhat}{\hat{H}}
\newcommand{\Uhat}{\hat{U}}

\newcommand{\Vhat}{\hat{V}}
\newcommand{\Mhat}{\hat{M}}
\newcommand{\Qj}{\hat{Q}_j}
\newcommand{\Qk}{\hat{Q}_k}
\newcommand{\Qjk}{\hat{Q}_{jk}}
\newcommand{\Qkj}{\hat{Q}_{kj}}
\newcommand{\Pjk}{\hat{P}_{jk}}
\newcommand{\Pk}{\hat{P}_k}

\newcommand{\mapV}{{\cal V}}
\newcommand{\mapU}{{\cal U}}
\newcommand{\mapM}{{\cal M}}
\newcommand{\Pfour}{{\cal P}_4}
\newcommand{\Qfour}{{\cal Q}_4}

\newcommand{\ketg}{\ket{\psi_1}}
\newcommand{\kets}{\ket{\psi_2}}
\newcommand{\kett}{\ket{\psi_3}}
\newcommand{\ketS}{\ket{\psi_4}}

\newcommand{\mutt}{\mu_{32}}
\newcommand{\tfo}{t_{41}}
\newcommand{\dt}{\delta t}
\newcommand{\Dt}{\Delta t}

\begin{document}

\title{Coherent chemical kinetics as quantum walks II: Radical-pair reactions in {\em Arabidopsis thaliana}}

\author{A. Chia}
\affiliation{Centre for Quantum Technologies, National University of Singapore}

\author{A. G\'orecka}
\affiliation{Division of Physics and Applied Physics, School of Physical and Mathematical Sciences, Nanyang Technological University, Singapore}

\author{P. Kurzy\'nski}
\affiliation{Centre for Quantum Technologies, National University of Singapore}
\affiliation{Faculty of Physics, Adam Mickiewicz University}

\author{T. Paterek}
\affiliation{Centre for Quantum Technologies, National University of Singapore}
\affiliation{Division of Physics and Applied Physics, School of Physical and Mathematical Sciences, Nanyang Technological University, Singapore}

\author{D. Kaszlikowski}
\affiliation{Centre for Quantum Technologies, National University of Singapore}

\date{\today}

\begin{abstract}

We apply the quantum-walk approach recently proposed in arXiv:quant-ph-1506.04213 to a radical-pair reaction where realistic estimates for the intermediate transition rates are available. The well-known average hitting time from quantum walks can be adopted as a measure of how quickly the reaction occurs and we calculate this for varying degrees of dephasing in the radical pair. The time for the radical pair to react to a product is found to be independent of the amount of dephasing introduced, even in the limit of no dephasing where the transient population dynamics exhibit strong coherent oscillations. This can be seen to arise from the existence of a rate-limiting step in the reaction and we argue that in such examples, a purely classical model based on rate equations can be used for estimating the timescale of the reaction but not necessarily its population dynamics.

\end{abstract}


\maketitle

\section{Introduction}
\label{Intro}

\begin{figure}[t]
\centerline{\includegraphics[width=7.5cm]{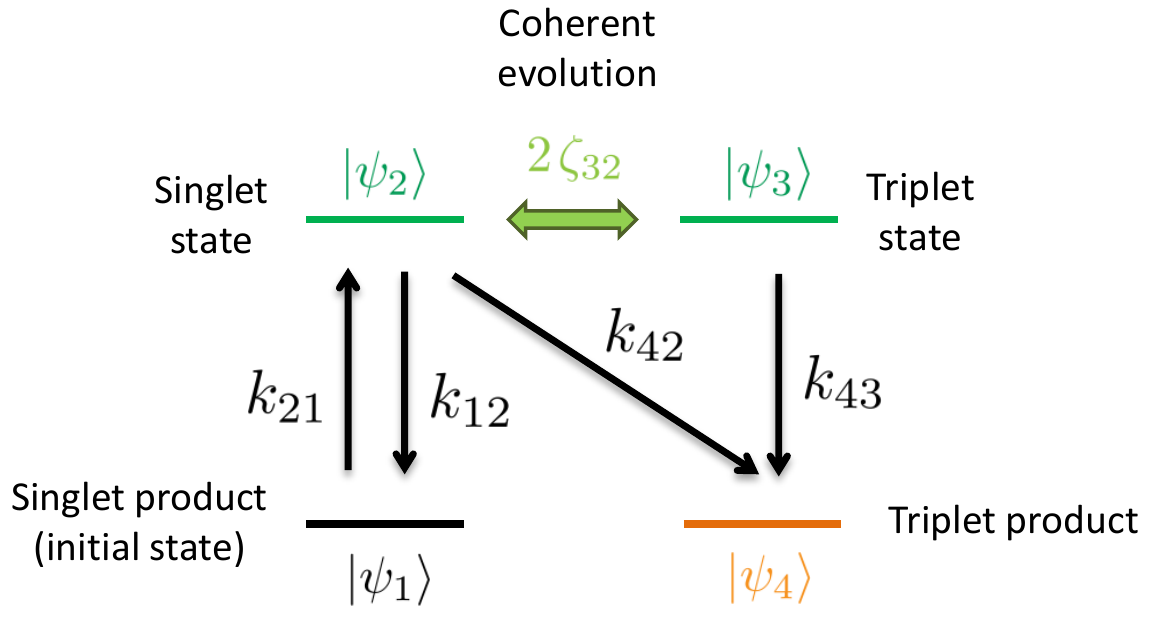}}
\caption{\label{f3} Schematic of the radical-pair reaction. The radical pair is assumed to be created in the singlet state $\ket{\psi_2}$ at a rate of $k_{21}$ from some precursor state of cryptochrome labelled as $\ket{\psi_1}$. The reaction can then proceed by having the radical pair decay back to the initial state or convert to a spin triplet state. The singlet-triplet interconversion is a coherent process occurring with rate $2\zeta_{32}$ which can be modulated by an applied magnetic field. As explained in the main text, the outcome of this reaction is amenable to an external magnetic field by having the path to the initial state open only to the the singlet state. Because $\ket{\psi_1}$ is also the result of the reaction associated exclusively to the singlet state, we will also refer to it as the singlet product. In line with conventional nomenclature we will refer to $\ket{\psi_4}$ as the ``triplet product'' although here this would be a misnomer because $\ket{\psi_4}$ is not uniquely associated with the triplet state.}
\end{figure}

It has previously been argued that the evolution of populations and coherences in what is known as the radical-pair reaction \cite{RH09} may be treated phenomenologically using the theory of quantum walks \cite{CGT+15}. The present paper is a continuation of Ref.~\cite{CGT+15} and we shall henceforth refer to Ref.~\cite{CGT+15} as ``Part I''. In essence Part I develops an approach to chemical reactions which takes the intermediate transition rates as inputs to the model, akin to classical rate equations except with coherences between different sites of the reaction taken into account. Here we apply this approach to an example where realistic estimates for the intermediate transition rates are available. The example is again a radical-pair reaction, shown in Fig.~\ref{f3}. This is a variant of the reaction originally proposed by Ritz and coworkers in Ref.~\cite{RAS00} and which was studied in Part I. The reaction of Fig.~\ref{f3} can be understood as an approximation of a real reaction (Fig.~\ref{f1} in Appendix~\ref{CryReact}) where certain fast transitions have been ignored. The actual reaction of Fig.~\ref{f1} is thought to occur in the cryptochromes of the plant \emph{Arabidopsis thaliana} which has been studied in Ref.~\cite{SCS07} and we will take realistic estimates for the various intermediate transition rates from there.
The reaction scheme of Fig.~\ref{f3} has been reviewed in Ref.~\cite{RH09} and its ability to act as a magnetic compass is supported by experimental evidence \cite{MRH+12} (see Fig.~1 of Ref.~\cite{RH09} and Fig.4 of Ref.~\cite{MRH+12}). As with the radical-pair mechanism of Part I, changes in the applied magnetic field are reflected in the product yields of the reaction. We will not be studying how the product yields depend on changes in an applied magnetic field. Instead we will study the effect of coherence on the reaction kinetics for a constant magnetic field with an Earth-strength intensity (approximately 50~$\mu$T).

We now briefly run through the basic principle of operation for the radical-pair reaction in Fig.~\ref{f3}: The radical pair is assumed to be created in a spin singlet state $\ket{\psi_2}$ from some initial state of cryptochrome $\ket{\psi_1}$ at a rate of $k_{21}$. The singlet state can then 1) decay to the singlet product (i.e.~return to the initial state) with rate $k_{12}$, 2) decay to the triplet product $\ket{\psi_4}$ at rate $k_{42}$, or 3) convert coherently to the spin triplet state $\ket{\psi_3}$ at rate $2\zeta_{32}$. The physical origin of the coherent interconversion between $\ket{\psi_2}$ and $\ket{\psi_3}$ comes from the hyperfine interactions internal to the radical-pair system. Under the application of an external magnetic field this coherent switching can be modulated by the Zeeman interaction which is sensitive to both the direction and magnitude of the applied field. Note the transition to the singlet product is spin-forbidden from the triplet state, only the singlet state is allowed to recombine to the singlet product but the triplet product can be reached from either the singlet or triplet state. This means that a magnetic field which enhances the coherent conversion of the radical pair from its singlet state to the triplet state will tend to increase the triplet yield (the amount of triplet products) whereas a magnetic field that reduces the singlet-triplet coherence tends to diminish the triplet yield. This renders the triplet yield (and hence the overall reaction) sensitive to an applied magnetic field. This means the various rates in Fig.~\ref{f3} are in general functions of the applied field and the rates that we will use in this paper correspond to a magnetic field in a given direction and intensity.

In Part I we focused mainly on the recombination process of the radical-pair reaction and its decoherent effect on the spin coherence of the radical pair (recall that the recombination process is the mechanism responsible for turning the radical pair into the products). Although modelling additional decoherent processes such as the g-anisotropy of Ref.~\cite{MLGH13} was considered (see Sec.~V.~A of Part I), its actual effect on the chemical reaction was never studied in detail. It is the intention of the present paper to study the effect of dephasing in the radical-pair reaction by using the dephasing map introduced in Part I. However, we do not attribute the dephasing to any physical mechanism, instead, we shall consider the dephasing strength to be a variable that we can tune. This allows us to study the quantum and classical limits of the reaction in the presence of recombination. The quantum limit then corresponds to setting the dephasing strength to zero, while the classical limit corresponds to setting the dephasing strength to its maximum value. We will find the radical-pair population to exhibit oscillations characteristic of coherent quantum evolution (or Rabi oscillations) for low enough dephasing strengths and that this oscillation becomes weaker as we increase the amount of dephasing. This allows us to judge whether a classical rate-equation model is sufficient for capturing the population dynamics, or if a quantum model is really necessary for a given set of intermediate transition rates and dephasing strength. However, the radical-pair population is not the only quantity that is sensitive to coherences. It is well known from quantum-walk theory that the time for the walk to reach a preassigned state also depends on the amount of coherence one can establish between different sites in the walk. This time is known as the hitting time \cite{Kem03a,Kem03b} (also known as the time of first passage in stochastic processes \cite{Red01}) and we will also look at how this changes as we vary the amount of dephasing in the radical pair. In the context of a chemical reaction this time can be taken as a measure of the time required for the reaction to happen and we find this to be essentially independent of the coherence in the radical pair due to the presence of a rate-limiting step \cite{ER14}. This means that the hitting time is only an interesting quantity to consider in the absence of such a rate-limiting step and we will suggest a problem in which this is the case in the conclusion of our paper.

The rest of the paper is organized as follows. The necessary tools for constructing a quantum-walk model of Fig.~\ref{f3} are covered in Sec.~\ref{QuantumWalk}. These results have already been covered in detail in Part I so here we will only summarise the key elements used in our simulation. These are the definitions of the so-called Kraus maps for amplitude damping, dephasing, and coherent evolution. These are then used in Sec.~\ref{RPMasQW} to construct a time-evolution map for the reaction. We will also introduce the concept of an average hitting time and calculate this in terms of the time-evolution map. This then allows us to associate the reaction time with the average hitting time of our quantum-walk model. We then simulate the radical-pair reaction and calculate its average hitting time by using rates obtained from Ref.~\cite{SCS07} for different dephasing strengths in Sec.~\ref{HittingTime}. We then conclude our paper in Sec.~\ref{Discussion} with a summary of our key results and mention a possible path for future exploration.

\section{Kraus maps for quantum walk}
\label{QuantumWalk}

\begin{figure}
\center
\includegraphics[width=6.5cm]{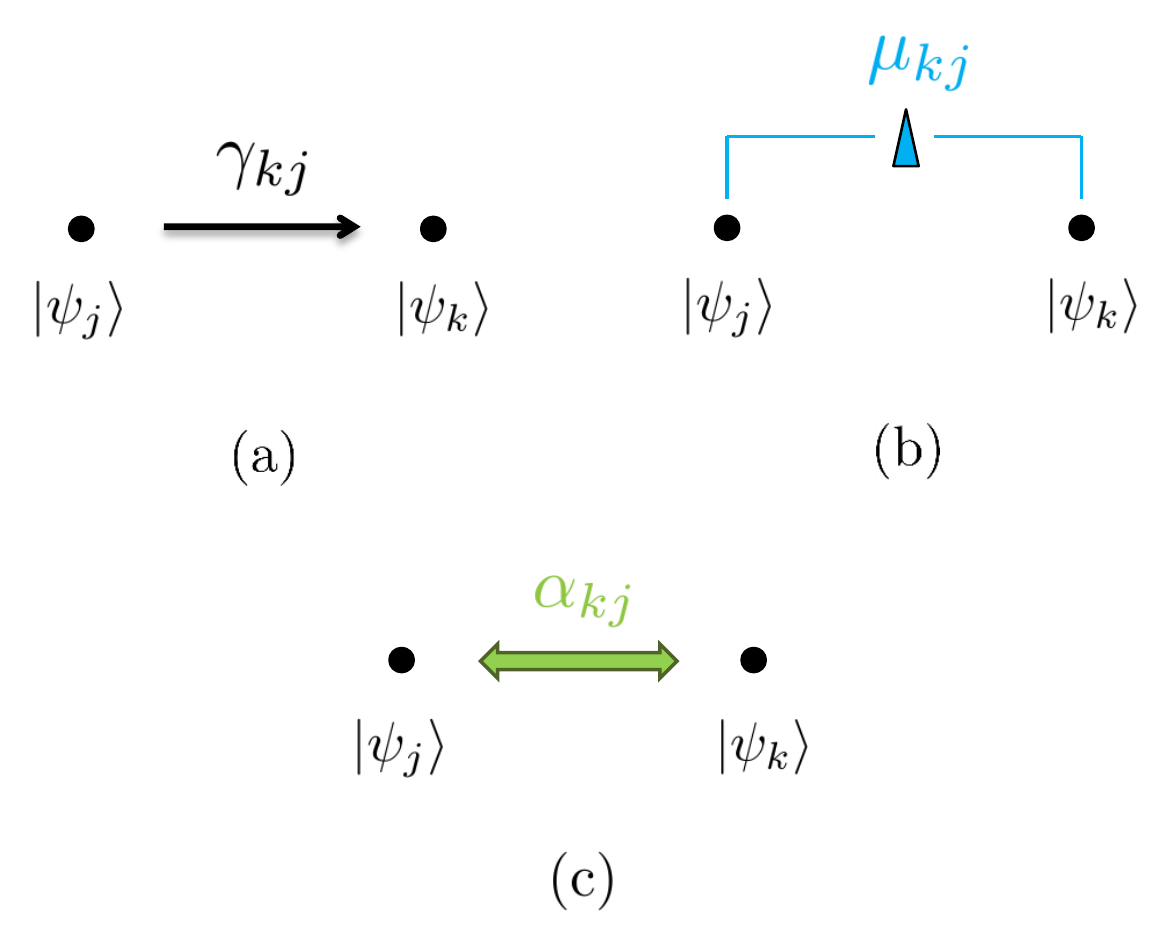} 
\caption{Depiction of the elementary processes used to simulate the quantum walk of Fig.~\ref{ChemReactGraph}. The states $\ket{\psi_j}$ and $\ket{\psi_k}$ are assumed to be any two states among an arbitrary number of states. (a) Amplitude damping from $\ket{\psi_j}$ to $\ket{\psi_k}$. This is represented by a one-way arrow which should remind us that this is an irreversible process. (b) Dephasing of states $\ket{\psi_j}$ and $\ket{\psi_k}$. The line represents coherence between $\ket{\psi_j}$ and $\ket{\psi_k}$ which has a wedge (represented by the triangle) driven into it, thereby destroying the ``connectedness'' of the two states. (c) Coherent oscillations between states $\ket{\psi_j}$ and $\ket{\psi_k}$. This is represented by a single two-way arrow rather than two one-way arrows to emphasise that this is a quantum coherent process.}
\label{BasicProcesses}
\end{figure}
The quantum-walk formalism visualises state transitions in a quantum system as a network of nodes (representing states) connected by edges (representing transitions), called graphs. Such models have a wide applicability because the nodes can represent abstract degrees of freedom, such as the different chemical compositions of molecules in a chemical reaction. We would therefore describe the reaction outlined in Fig.~\ref{f3} by simply representing the different chemical states as nodes on a graph. Each node is labelled by a state $\ket{\psi_k}$ with the value of $k$ consistent with Fig.~\ref{f3}. The corresponding quantum-walk model of Fig.~\ref{f3} can then be constructed by using the interconnections shown in Fig.~\ref{BasicProcesses}. The final graph corresponding to Fig.~\ref{f3} is shown in Fig.~\ref{ChemReactGraph}. Each interconnection in Fig.~\ref{BasicProcesses} is defined by a Kraus map and the goal of this section is to first go through what these are as they will be used in the next section to describe the full quantum walk in Fig.~\ref{ChemReactGraph}. As we have already treated these interconnections in detail in Part I, this section on Kraus maps is only meant to be a recapitulation. A reader familiar with Kraus maps or have read Part I in detail may wish to proceed directly to Sec.~\ref{RPMasQW} from here.

\subsection{Amplitude damping}
\label{AmpDamping}

The incoherent transfer of population from one state $\ket{\psi_j}$ to another $\ket{\psi_k}$, as symbolised by Fig.~\ref{BasicProcesses}~(a), can be accomplished by the following Kraus map:
\begin{align}
\label{Mjk}
	\mapM_{jk}(\Dt) \, \rho(t) = {}& \Mhat^{(1)}_{jk}(\Dt) \, \rho(t) \, \Mhat^{(1)}_{jk}{}\dg(\Dt)  \nn \\
	                               & + \Mhat^{(2)}_{jk}(\Dt) \, \rho(t) \, \Mhat^{(2)}_{jk}{}\dg(\Dt)  \;,
\end{align}
with the Kraus operators
\begin{gather}
\label{M1}
	\Mhat^{(1)}_{jk}(\Dt) = \sqrt{\gamma_{jk}(\Dt)} \: \Qjk \;, \\
\label{M2}
	\Mhat^{(2)}_{jk}(\Dt) = \Pk + \sqrt{1-\gamma_{jk}(\Dt)} \: \Qk \;,  
\end{gather}
where $\gamma_{jk}(\Dt) \in [0,1]$ and we have defined 
\begin{gather}
	\Qjk = \op{\psi_j}{\psi_k}  \;,  \\
	\Qk = \op{\psi_k}{\psi_k} \;,  \quad  \Pk = \hat{1} - \Qk   \;.  
\end{gather}

\begin{figure}[t]
\center
\includegraphics[width=4cm]{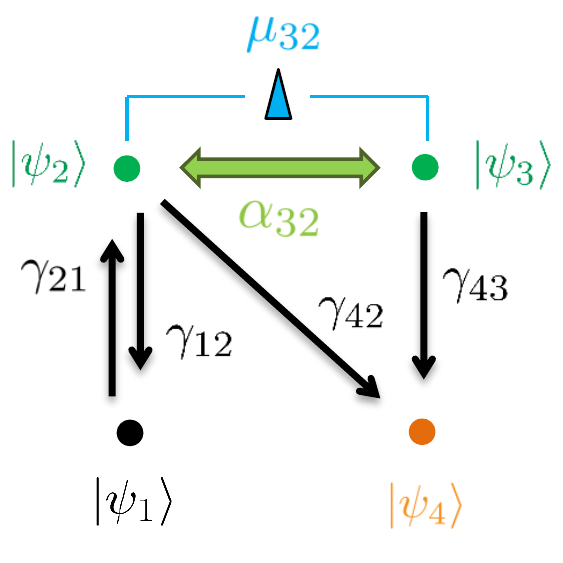} 
\caption{Representation of the chemical reaction in Fig.~\ref{f3} as a quantum walk. The interconnections between nodes are shown individually in Fig.~\ref{BasicProcesses} and described in the main text.}
\label{ChemReactGraph}
\end{figure}

The dimensionality of $\rho(t)$ in \eqref{Mjk}--\eqref{M2} is arbitrary so that \eqref{Mjk} applies to any two states $\ket{\psi_j}$ and $\ket{\psi_k}$ out of an arbitrary number of states (although for Fig.~\ref{ChemReactGraph} we have only four states). The map is characterised by the probability of a transition from $\ket{\psi_j}$ to $\ket{\psi_k}$ over the interval $\Dt$ which can be expressed in terms of the rate of transition $k_{ij}$ as 
\begin{align}
\label{ADRate}
	\gamma_{ij} = k_{ij} \; \Delta t  \;.
\end{align}
Realistic estimates of $k_{ij}$ for different $i$ and $j$ will be taken from Ref.~\cite{SCS07}.

\subsection{Dephasing}
\label{CohEvoDeph}

A map which removes only the coherences between $\ket{\psi_j}$ and $\ket{\psi_k}$ while leaving their populations untouched has the Kraus form
\begin{align}
\label{Vjk}
	{\cal V}_{jk}(\Dt) \, \rho(t) = {}& \Vhat^{(1)}_{jk}(\Dt) \, \rho(t) \, \Vhat^{(1)}_{jk}{}\dg(\Dt)  \nn  \\
	                                  & + \Vhat^{(2)}_{jk}(\Dt) \, \rho(t) \, \Vhat^{(2)}_{jk}{}\dg(\Dt)  \;,
\end{align}
where 
\begin{gather}
\label{V1}
	\Vhat^{(1)}_{jk}(\Dt) = \sqrt{\mu_{jk}(\Dt)} \; \Qk  \;, \\
\label{V2}
	\Vhat^{(2)}_{jk}(\Dt) = \Pk + \sqrt{1-\mu_{jk}(\Dt)} \; \Qk \;.   
\end{gather}
As with amplitude damping, we can work with the rate of dephasing rather than with probability $\mu_{jk}$. Denoting the rate of dephasing between states $\ket{\psi_j}$ and $\ket{\psi_k}$ as $q_{jk}$, we can write
\begin{align}
\label{DecParam}
	\mu_{jk} = q_{jk} \, \Delta t  \;.
\end{align}
The purpose of introducing dephasing is to allow for a variable amount of decoherence in the system. When we simulate the reaction of Fig.~\ref{ChemReactGraph} we will set $\Delta t$ to some small number $\dt$ and change $q_{jk}$. The restriction of $\mu_{jk}$ to be between zero and one then implies that $q_{jk} \in [0,1/\dt]$. We would then like to obtain the classical limit (represented by a $\rho$ with all off-diagonal elements equal to zero at all times $t_n=n\,\dt$) by setting $q_{jk}=1/\dt$. Note that there is only one decoherence parameter present in Fig.~\ref{ChemReactGraph}, given by $\mutt$ so we will only have $q_{32}$ to vary at will (as far as dephasing is concerned).

\subsection{Coherent evolution}

Coherent oscillations between states $\ket{\psi_j}$ and $\ket{\psi_k}$ can be captured by unitary evolution
\begin{align}
\label{Ujk}
	{\cal U}_{jk}(\Dt) \, \rho(t) = \Uhat_{jk}(\Dt) \, \rho(t) \, \Uhat\dg_{jk}(\Dt)  \;,
\end{align}
where 
\begin{align}
	\Uhat_{jk}(\Dt) = e^{-i \Hhat_{jk} \Dt}  \;,
\end{align}
(with $\hbar \equiv 1$ for convenience) and for $j \ne k$,
\begin{align}
\label{Hhatjk}
	\Hhat_{jk} = \omega_j \, \Qj + \omega_k \, \Qk + \Omega_{jk} \big( \, \Qjk + \Qkj \, \big)  \;.
\end{align}
Here $\omega_k$ is the expectation value of $\Hhat_{jk}$ in the state $\ket{\psi_k}$ while the coupling between states $\ket{\psi_j}$ and $\ket{\psi_k}$ is denoted by $\Omega_{jk}$. We assume that $\Omega_{jk}$ to be real and symmetric with respect to its indices so that $\Hhat_{jk}$ is Hermitian. The map \eqref{Ujk} can also be parameterized by the probability of a transition that it induces, which is given by 
\begin{align}
	\alpha_{jk}(\Dt) \equiv {}& {\rm Pr}\big[ \rho(t) =  \Qj  \, \big| \, \rho(0) = \Qk \big]  \nn \\
                        = {}& \big| \bra{\psi_j} \Uhat_{jk}(t) \ket{\psi_k} \big|^2  \;.
\end{align}
We have used the notation ${\rm Pr}[A|B]$ to denote the probability of event $A$ occurring given the occurrence of event $B$. We will evaluate the unitary operator in closed form in Appendices~\ref{AppA} and \ref{AppB} which in turn gives us an expression for the transition probability:
\begin{align}
\label{PrQjQk}
	\alpha_{jk}(\Dt) = \frac{\Omega^2_{jk}}{2\,\zeta^2_{jk}} \; \big[ 1 - \cos\big( 2\;\!\zeta_{jk}\;\! t \big) \big]  \;,
\end{align}
where
\begin{align}
\label{ParameterDefnB}
	\zeta_{jk} = \frac{1}{2} \, \rt{(\omega_k - \omega_j)^2 + 4\Omega^2_{jk}}  \;.
\end{align}
The frequency at which the system oscillates between states $\ket{\psi_j}$ and $\ket{\psi_k}$ can then be defined to be the frequency at which \eqref{PrQjQk} oscillates, which is $2\,\zeta_{jk}$. From \eqref{PrQjQk} and \eqref{ParameterDefnB} we see that increasing $|\omega_j-\omega_k|$ will lower the transition probability between $\ket{\psi_j}$ and $\ket{\psi_k}$ while increasing its frequency of oscillation.

\section{Radical-pair reaction as a quantum walk}
\label{RPMasQW}

\subsection{Time evolution map}

For the purpose of describing the radical-pair reaction the set $\{\ket{\psi_k}\}_{k=1}^4$ can be taken to be complete and spans the system Hilbert space so that
\begin{align}
\label{IdRes}
	\sum_{k=1}^{4} \; \op{\psi_k}{\psi_k} =	\hat{1}   \;.
\end{align}
We will then represent an arbitrary time-dependent state $\rho(t)$ by a $4 \times 4$ matrix in the basis $\{\ket{\psi_k}\}_{k=1}^4$. The evolution of $\rho(t)$ over any finite time $\Delta t$ can be described generally by a superoperator $\mapK(t+\Delta t,t)$. Since the evolution over a finite interval can always be obtained by composing infinitely many infinitesimal time steps we will only consider the case when $\Delta t$ is small and work in discrete time. Of course, if we want to simulate the evolution of $\rho(t)$ on a computer we will have to discretise time. In this case we can never have a true infinitesimal time step $dt$, but as long as our time steps are sufficiently small the true dynamics of $\rho(t)$ in continuous time will be well approximated by its discrete-time version. Here we will use $\dt$ to denote a small but finite time step to distinguish it from a true infinitesimal. For sufficiently small $\dt$ we can parameterize $\mapK$ by a single time argument and we write
\begin{align}
\label{rho(tn)}
	\rho(t_n) = \big[ \;\! \mapK(\dt) \big]^n \, \rho(t_0)  \;,
\end{align}	
where $t_n \equiv t_0 + n \,\dt$ with $n$ being any non-negative integer. A suitable choice for $\dt$ depends on the various intermediate transition rates of the quantum walk. We discuss how an appropriate value of $\dt$ is determined in Appendix~\ref{CryReact} in conjunction with a discussion of how the various transition rates are chosen.

All the dynamics taking the initial state $\ketg$ to the triplet product $\ketS$ is now encapsulated in the map $\mapK(\dt)$. To determine its form we can simply read off Fig.~\ref{ChemReactGraph} where each transition can be ascribed to one of the processes defined in Sec.~\ref{QuantumWalk}. This gives
\begin{align}
\label{Kb}
	\mapK(\dt) = {}& \mapM_{43}(\dt) \, \mapM_{42}(\dt) \, \mapV_{32}(\dt) \, \mapU_{32}(\dt) \nn \\
	               & \times  \mapM_{12}(\dt) \, \mapM_{21}(\dt) \;.
\end{align}
We make a few important remarks on our quantum-walk model in the following.
\begin{enumerate}

\item Following from Part I we see that as long as $\dt$ is small the propagation of $\rho(0)$ according to \eqref{rho(tn)} will be insensitive to the order of the various maps in \eqref{Kb}. In discrete time this can be proven rigorously using the Lie-Trotter formula \cite{RH12}. This is analogous to using what is known as the Zassenhaus formula in the case of pure states and neglecting terms on the order of $\dt^2$ \cite{JKP12}. This means that $\dt$ should be small compared to all the rates in the system. 

\item All coherences (off-diagonal elements in $\rho$) are zero except for $\rho_{32}$ and $\rho_{23}$ for any initial state which is diagonal. This is because only $\mapU_{32}(\dt)$ appears in \eqref{Kb} and this is the only part in $\mapK(\dt)$ that is capable of creating coherences beween states corresponding to its two subscripts. Although the maps $\mapM_{jk}(\dt)$ and $\mapV_{jk}(\dt)$ do change the coherences between states $\ket{\psi_j}$ and $\ket{\psi_k}$, they will only diminish it. So if there are no coherences between $\ket{\psi_j}$ and $\ket{\psi_k}$ (which is the case for an initial state that is diagonal) then $\mapM_{jk}(\dt)$ changes only their populations while $\mapV_{jk}(\dt)$ has no effect. As we will be assuming that our initial state is $\ketg$, we have only one number which characterises the coherences in the system, namely $\rho_{32}$ (and we know that $\rho$ is Hermitian so $\rho_{32}=\rho^*_{23}$).

\item  We should also mention that the triplet product in Fig.~\ref{f3} is often referred to as a signalling state, related to how molecules can ``communicate'' with each other via a process known as signalling in biology \cite{DHW13,CPB+11,BWGS14}. Although we will not be interested in signalling process itself, we should note that it is essential for cryptochrome to be in this state in order to participate in magnetoreception, i.e.~cryptochrome is considered to be ``active'' when it is in this state and its signalling activity during magnetoreception can be measured by the yield of this state \cite{SCS07}. For this reason we consider the reaction modelled by Fig.~\ref{ChemReactGraph} to be complete when $\ket{\psi_4}$ (corresponding to the triplet product state) is reached. The time for the radical-pair reaction to happen will thus be measured by the time it takes the quantum walk to go from $\ket{\psi_1}$ to $\ket{\psi_4}$.

\end{enumerate}

\subsection{Hitting-time distribution and the average hitting time}

The hitting time of a random walk (classical or quantum) is the time taken to reach a preassigned state for the first time from a given initial state in one specific realization of the walk. The average hitting time is then the average of hitting times obtained over many realizations of the random walk. Here we define the average hitting time and calculate an expression for it in terms of the evolution map \eqref{Kb}. A reason for considering the average hitting time is that it is a function of the coherences in the quantum walk. Thus in general the average hitting time for a quantum walk will be different to a classical random walk where there are no coherences.

In discrete time the average hitting time can be characterised by the average number of steps taken to reach $\ketS$ starting from $\ketg$. If we denote the number of steps taken to reach $\ketS$ by $n$, and its probability distribution by $f_{41}(n)$, the average value of $n$ is then defined by
\begin{align}
\label{n41Sum}
	n_{41} = \sum_{n=0}^{\infty} \; n \, f_{41}(n)  \;.
\end{align}
The average hitting time will simply be
\begin{align}
\label{t41}
	\tfo = n_{41} \, \dt \;. 
\end{align}
We will loosely refer to $f_{41}(n)$ as the hitting-time distribution even though it is actually the probability for the number of steps to reach $\ketS$. To calculate $n_{41}$ we first need to find $f_{41}(n)$, which is defined as
\begin{align}
\label{f41Defn}
	f_{41}(n) \equiv {}& {\rm Pr}\big[ \rho(t_n) = \hat{Q}_4 \, | \, \rho(0) = \hat{Q}_1 ,  \nn \\
	                   & \rho(t_m) \ne \hat{Q}_4  \;\forall \; m \le n-1 \big] \;.
\end{align}
An important difference between quantum and classical hitting-time distributions lies in the fact that a quantum system can be in a coherent superposition of states whereas a classical system cannot. This means that we have to measure a quantum system to see if it is in a particular state or not. For this reason the conditional probability \eqref{f41Defn} has to refer to a sequence of measurements which expresses the knowledge that the system is not in state $\ketS$ for all times prior to $t_n$. Since we are only interested in whether the system is in $\ketS$ or not at each time, the measurement outcome is binary. The change brought upon the system state by such a measurement can again be effected by Kraus maps in the following manner: If the system state is $\rho(t_n)$ before the measurement, its state immediately after the measurement given that it is found in $\ketS$ is
\begin{align}
\label{rho1}
	\rho_1(t_n) \equiv \frac{ \hat{Q}_4 \, \rho(t_n) \, \hat{Q}_4 }{ {\rm Tr}\big[ \hat{Q}_4 \, \rho(t_n) \, \hat{Q}_4 \big] }  \;,
\end{align}
where the denominator in \eqref{rho1} normalises $\rho_1(t_n)$. It is simply the probability of finding the system in state $\ketS$
\begin{align}
	{\rm Pr}\big[ \rho(t_n)=\hat{Q}_4 \big] = {\rm Tr}\big[ \hat{Q}_4 \, \rho(t_n) \, \hat{Q}_4 \big]  
	                                        = \bra{\psi_4} \rho(t_n) \ket{\psi_4} \;,
\end{align}
which is just its occupation probability at time $t_n$. We will call a measurement which reveals the system to not be in state $\ketS$ a null measurement. The system state immediately after a null measurement is given by
\begin{align}
	\rho_0(t_n) \equiv \frac{ \hat{P}_4 \, \rho(t_n) \, \hat{P}_4 }{ {\rm Tr}\big[ \hat{P}_4 \, \rho(t_n) \, \hat{P}_4 \big] }  \;.
\end{align}
Since the measurement has only two possible outcomes the probability of not finding the system in state $\ketS$ is simply 
\begin{align}
	{\rm Pr}\big[ \rho(t_n)\ne\hat{Q}_4 \big] = {\rm Tr}\big[ \hat{P}_4 \, \rho(t_n) \, \hat{P}_4 \big]  
	                                          = 1 - {\rm Pr}\big[ \rho(t_n)=\hat{Q}_4 \big] .
\end{align}
Following this prescription we can express the conditioning in \eqref{f41Defn} as a sequence of null measurements (applications of $\hat{P}_4)$ at times $t_m$ for $m \le n-1$, each separated by $\mapK(\dt)$. For notational convenience we define the maps
\begin{align}
	\Pfour \, \rho = \hat{P}_4 \, \rho \, \hat{P}_4  \;, \quad  \Qfour \, \rho = \hat{Q}_4 \, \rho \, \hat{Q}_4  \;.
\end{align}
The hitting-time distribution as defined by \eqref{f41Defn} is then given by
\begin{align}
\label{f41Final}
	f_{41}(n) = {\rm Tr}\!\left\{ \Qfour \, \mapK(\dt) \big[ \Pfour \, \mapK(\dt) \, \big]^{n-1} \rho(0) \right\}  \;.
\end{align}
Note that $n \ge 1$ in this expression. For $n=0$ we have $f_{41}(0)=0$ because the process begins at $\rho(0)=\op{\psi_1}{\psi_1}$. In principle we are done since $n_{41}$ is just the weighted sum \eqref{n41Sum} with $f_{41}(n)$ given by \eqref{f41Final}. However we can proceed further by noting that the statistical moments of a probability distribution can also be derived from the distribution's generating function \cite{KMT12}. It is simple to show that the first moment of $f_{41}(n)$, i.e.~the mean of $n$, is related to its generating function $F_{41}(z)$ by
\begin{align}
\label{dF41/dz}
	n_{41} = \bigg[ \frac{d}{dz} \, F_{41}(z) \bigg|_{z=1}  \;,
\end{align}
where $F_{41}(z)$ is defined by the power series
\begin{align}
\label{F41}
	F_{41}(z) = \sum_{n=0}^{\infty} \, f_{41}(n) \, z^n  \;.
\end{align}
Details of the derivation of $n_{41}$ can be found in Appendix~\ref{DerivationOfn41}. A similar calculation can also be found in Ref.~\cite{KB06} but for a system following unitary evolution. The result of substituting \eqref{f41Final} and \eqref{F41} in \eqref{dF41/dz} and simplifying is  
\begin{align}
\label{n41Trace}
	n_{41} = {\rm Tr}\Big\{ \Qfour \, \mapK(\dt) \Big( \big[ \mathbbm{1} - \Pfour \, \mapK(\dt) \big]^{-1} \Big)^2 \rho(0) \Big\} \;. 
\end{align}
Let us note a few points regarding this expression. The first is that \eqref{n41Trace} is an exact formula for $n_{41}$. For numerical evaluations \eqref{n41Trace} is simpler to use compared to the weighted sum \eqref{n41Sum} because \eqref{n41Sum} has to be truncated at some $n$. Such a value of $n$ is determined from the normalisation of $f_{41}(n)$ and is permissible provided that $f_{41}(n)$ is effectively normalised. On the other hand \eqref{n41Trace} can be used without the need to preexamine $f_{41}(n)$. When the hitting-time distribution has a long tail it is also faster to use \eqref{n41Trace} compared to \eqref{n41Sum}. Second, $n_{41}$ depends on the size of $\dt$ since this is the average number of steps required for the system to reach $\ketS$ for the first time. The smaller the step size the more steps the system must take to get to $\ketS$. The actual time however will not depend on $\dt$ provided that it is small enough. Finally, the definition of hitting time adopted here though sensible, is not universal. As alluded to under \eqref{f41Defn}, the ability for a quantum system to have a wavefunction spread over many sites makes the system's ``location'' a fuzzy concept. Because of this the quantum hitting time is not uniquely defined and it may be advantageous to use alternative definitions when more information about the quantum walk is available \cite{Kem03b}. In this paper we will only use \eqref{n41Trace}, which stems from the definition \eqref{f41Defn} for $f_{41}(n)$. In the next section we illustrate how \eqref{rho(tn)} [together with \eqref{Kb}], \eqref{f41Final}, and \eqref{n41Trace} behave for different values of $\mutt$ and for some suitably chosen set of system parameters.

\section{Simulation results of the radical-pair quantum walk}
\label{HittingTime}

\subsection{Results for realistic rates}
\label{CrypHittingTime}

Somewhat realistic estimates of the transition rates relevant to our simplified picture (Fig.~\ref{ChemReactGraph}) can be obtained from the literature. We explain how the rates of the various transitions in Fig.~\ref{ChemReactGraph} are identified with the estimates in Ref.~\cite{SCS07} in Sec.~2 of Appendix~\ref{CryReact}. The resulting rates are summarised in Table~\ref{RatesSummary} and are used in all the plots unless otherwise stated. We also obtain all our results by using the initial state ($t_0 \equiv 0$)
\begin{align}
	\rho(0) = \op{\psi_1}{\psi_1}  \;.
\end{align}
\begin{table}
	\begin{center}
		\begin{tabular}{c c}
		\hline
  	\hline
		Rate                & Value used in simulation (${\rm s}^{-1}$) \\    
    \hline 
    $k_{21}$            &   $1 \times 10^8$  \\ 
    $k_{12}$            &   $1 \times 10^{7}$  \\ 
    $k_{43},\,k_{42}$   &   $3.3 \times 10^6$  \\
		$\omega_3$          &   $1.76 \times 10^7$  \\
		$\omega_2$          &   $0$  \\
    $\Omega_{32}$       &   $4.06 \times 10^7$  \\
		$1/\dt$             &   $1 \times 10^{14}$  \\
		$q_{32}$            &   $[0, 10^{14}]$  \\
  	\hline
  	\hline
		\end{tabular}
	\end{center}
	\caption{\label{RatesSummary}Summary of rates used to simulate the graph of Fig.~\ref{ChemReactGraph}. The correspondence to rates in Fig.~\ref{f1} and Ref.~\cite{SCS07} are explained in Sec.~2 of Appendix~\ref{CryReact}.}
\end{table}

In Fig.~\ref{t41andtc}~(a) we plot the average hitting time $\tfo$ against $\mutt$ in log scale. Figure~\ref{t41andtc}~(a) was generated using \eqref{t41} and \eqref{n41Trace}. It shows that the average hitting time of cryptochrome is a monotonically increasing function of the decoherent noise in the system. Here we will use the terms decoherence and dephasing interchangeably when referring to $\mutt$. We discuss the qualitative features of the average hitting time first and the significance (or insignificance) of its numerical value in the next paragraph. As we increase the amount of decoherence in the radical-pair reaction, its average hitting time remains constant until $\mutt \approx 10^{-6}$ after which it becomes extremely sensitive to decoherence. This sensitivity to decoherent noise is only over a window of approximately three orders of magnitude in the noise strength ($10^{-6}$--$10^{-3}$). As we further increase the decoherent noise the average hitting time flattens off and becomes constant again. We thus find three distinctive segments in $\tfo$: 1) a quantum regime ($\mutt \longrightarrow 0$) for which the reaction time is robust to decoherence, 2) a classical regime ($\mutt \longrightarrow 1$) where the reaction time is also insensitive to decoherence, and 3) a quantum-to-classical transition where $\tfo$ rises sharply with increasing decoherence. The constancy of $\tfo$ in the classical limit can be understood by noting that the effect of a nonzero $\mutt$ is observable only if the system has nonzero coherences. Changes in the coherences of the system (i.e.~$\rho_{32}$) are then reflected (in a nontrivial way) in the changes of $\tfo$. However, coherences in the system vanish for $\mutt \approx 10^{-3}$ so increasing $\mutt$ beyond this point will not produce any changes in $\tfo$ (we show plots of coherences and occupation probabilities for selected amounts of decoherence in Fig.~\ref{rho}, which are discussed below). By the same token the constancy of $\tfo$ in the quantum limit arises because there, the decoherent noise is too weak to bring about any significant changes in the system coherences. The monotonicity of $\tfo$ as a function of $\mutt$ is thus attributed to the behaviour of the quantum-to-classical transition. This depends on the values of the system parameters which in turn determine the probability amplitudes for the different paths taken by cryptochrome to reach state $\ketS$ starting from $\ketg$ (recall Fig.~\ref{ChemReactGraph}). The precise way in which the different paths of the quantum walk interfere then gives rise to the quantum-to-classical segment seen in Fig.~\ref{t41andtc}. Therefore the statement that a more coherent (or quantum) system will react faster than a less coherent one is in fact not warranted in general, although it is true for the parameters used to plot Fig.~\ref{t41andtc}~(a). We illustrate this fact in Fig.~\ref{t41NonEstA} by using system parameters which deviate from those shown in Table~\ref{RatesSummary} (to be discussed later). We mentioned in Appendix~\ref{CryReact} that the value of $k_{21}$ adopted in Fig.~\ref{t41andtc}~(a) is inferred from a photolyase measurement, not cryptochrome, so we have also considered $\tfo$ by changing $k_{21}$ by an order of magnitude above and below $10^{8}\,{\rm s}^{-1}$. We have not shown these results but the curves are qualitatively the same as Fig.~\ref{t41andtc}~(a), retaining the step-like behaviour as a function of $\mutt$. For a fixed value of $\mutt$, increasing $k_{21}$ will increase the $\ketg \longrightarrow \kets$ transition probability which in turn decreases the average hitting time. Similarly, decreasing $k_{21}$ increases the average hitting time. Thus the $\tfo$ curve in Fig.~\ref{t41andtc}~(a) simply shifts down or up corresponding to an increase or decrease in $k_{21}$. 
\begin{figure}
\center
\includegraphics[width=7.5cm]{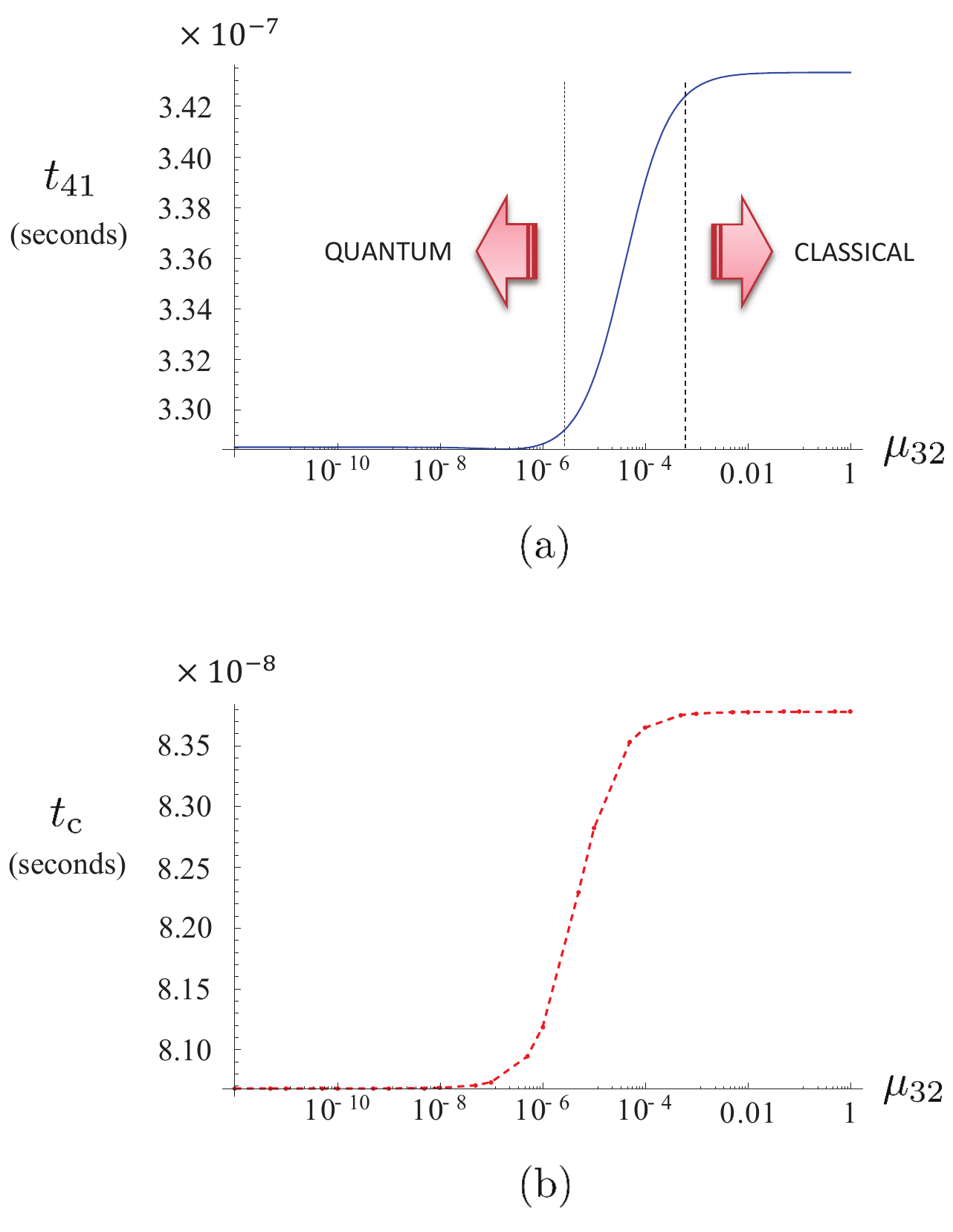} 
\caption{(a) Plot of the average hitting time as a function of the dephasing noise. The dephasing-noise axis is in log scale but the average hitting time is in linear scale. (b) Plot of the time it takes for the triplet-product state occupation probability to reach $0.2$ as a function of the dephasing noise. Both (a) and (b) are generated using the parameter values in Table~\ref{RatesSummary}.}
\label{t41andtc}
\end{figure}

The order of magnitude of $\tfo$ in Fig.~\ref{t41andtc}~(a) is consistent with the rates from Table~\ref{RatesSummary}. If we were to naively estimate the order of magnitude of $\tfo$ then one possibility is to regard the random walk as a classical process and add the times for each forward transition (ignoring the backward transitions for simplicity). That is, we would approximate $\tfo$ by $k^{-1}_{21}+(2\zeta_{32})^{-1}+k^{-1}_{42}$. From Table~\ref{RatesSummary} we can see that $k_{42} \approx 2\zeta_{32}/10 \approx k_{21}/100$ (i.e.~it is an order of magnitude less than the next highest rate) so the dominant term in our order-of-magnitude estimate is $k^{-1}_{42}$. Evaluating its inverse gives $k^{-1}_{42}=0.303 \, \mu{\rm s}$, which would be our ballpark figure for $\tfo$. Since we have ignored the backward transitions and $k^{-1}_{21}+(2\zeta_{32})^{-1}$ in our naive estimate the actual value of $\tfo$ should be a bit greater than $0.303 \, \mu{\rm s}$ and this is what we observe in Fig.~\ref{t41andtc}~(a). When we look at the actual numbers of the average hitting time in Fig.~\ref{t41andtc}~(a), we find the difference in $\tfo$ in going from $\mutt=1$ to $\mutt=0$ is only about four percent of the value at $\mutt=1$. This is due to the fact that the final transition to $\ket{\psi_4}$ from $\ket{\psi_2}$ and $\ket{\psi_3}$ is an incoherent process and the fact that these transitions have a significantly slower rate than other transitions in the reaction. This makes the $\ket{\psi_2} \longrightarrow \ket{\psi_4}$ and $\ket{\psi_3} \longrightarrow \ket{\psi_4}$ transtions examples of a rate-limiting step \cite{ER14}. One can therefore expect the reaction's timescale to be caputred by a purely classical rate-equation model. However this does not imply that the classical model can reproduce the correct population dynamics. This can be seen since in the limit of $\mutt \longrightarrow 0$ one can expect strong oscillations in $\rho_{22}$ and $\rho_{33}$, but $\tfo$ is only marginally different to its value when $\mutt \longrightarrow 1$. Therefore as a result of the rate-limiting step, we have an example reaction where a purely classical rate-equation model would suffice in describing the reaction's timescale even if it cannot capture the transient dynamics of the populations. The evolution of the populations is described later in Fig.~\ref{rho}.

\begin{figure}
\center
\includegraphics[width=8cm]{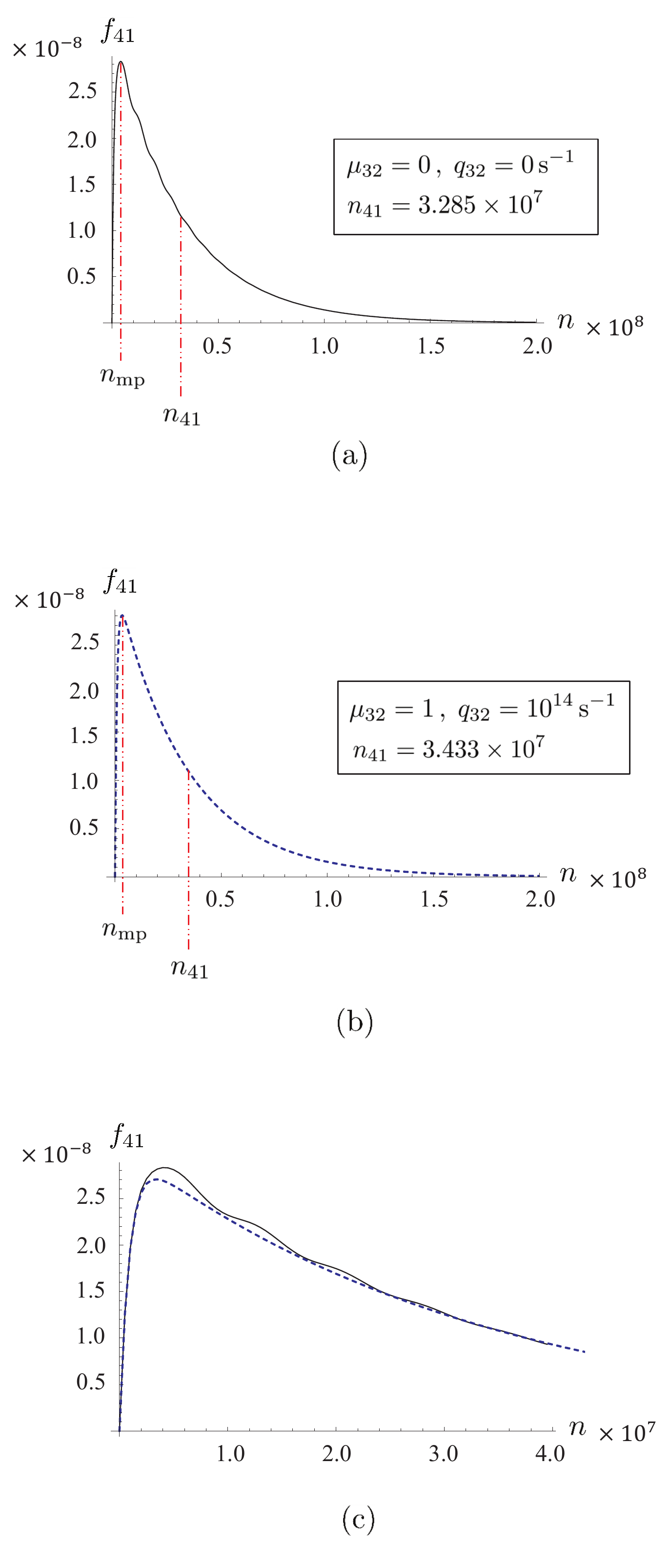} 
\caption{(a) The hitting-time distribution in the quantum limit. (b) The hitting-time distribution in the classical limit. (c) Close-up of the hitting-time distributions in (a) and (b) so that the oscillations in the distribution can be seen clearly. The average and the most probable values of $n$ for the quantum and classical limits have been marked by red dash-dot lines.}
\label{f41}
\end{figure}

\begin{figure*}
\center
\includegraphics[width=0.85\textwidth]{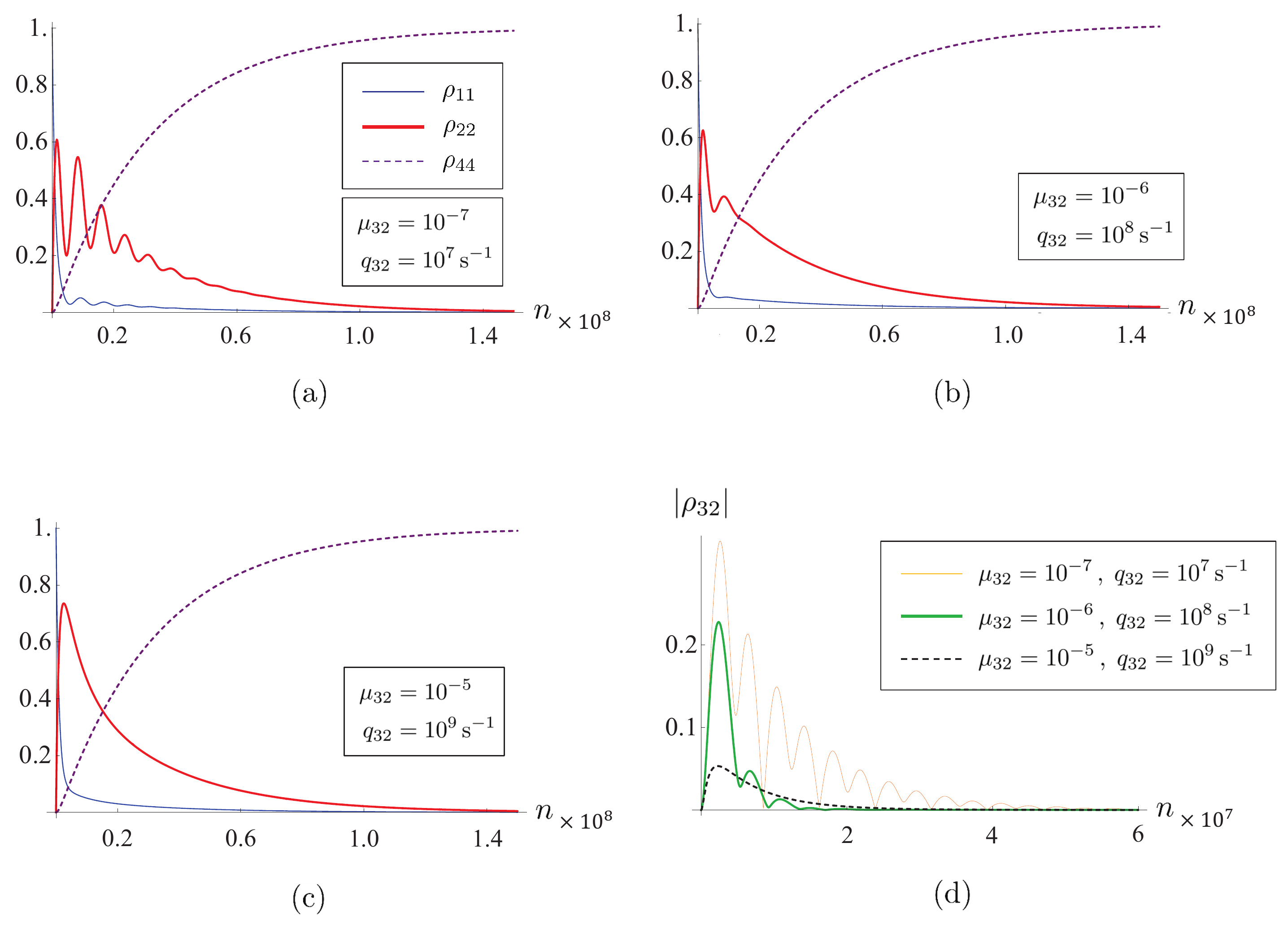} 
\caption{State-occupation probabilities $\rho_{11}$ (thin blue curve), $\rho_{22}$ (thick red curve), and $\rho_{44}$ (dashed purple curve) for (a) a relatively small amount of dephasing, (b) intermediate amount of dephasing, and (c) a large amount of dephasing. (d) The magnitude of coherences in the system for the different amounts of dephasing considered in (a)--(c) and shown in the inset.}
\label{rho}
\end{figure*}

Aside from the average hitting time, one can also characterise the reaction timescale by finding the time required for the occupation probability of the triplet product to reach some predefined value $\eta$. We define this time by $t_{\rm c}$, which is the solution to
\begin{align}
\label{tc}
	\rho_{44}(t_{\rm c}) = \eta 
\end{align}
for some $\eta \in [0,1]$. The time $t_{\rm c}$ is in fact simpler to calculate than the average hitting time as it only requires one to simulate the triplet product population. Figure~\ref{t41andtc}~(b) shows a plot of $t_{\rm c}$ for $\eta=0.2$. It can be seen from Fig.~\ref{t41andtc}~(b) that $t_{\rm c}$ is qualitatively the same as $\tfo$ except that it is less than $\tfo$ for every value of the dephasing parameter. To see why the $t_{\rm c}$ curve looks the same as the $\tfo$ curve let us consider a sample of a hundred cryptochrome molecules for the sake of argument. The value of $t_{\rm c}$ is then determined by the time it takes for the quickest twenty molecules to reach state $\ketS$ given that they all started at $\ketg$. But this time is determined by exactly when the twentieth molecule reaches state $\ketS$. If we were asked to estimate its arrival time for different values of dephasing, we would be guided by $\tfo$ but subtract a small amount from it. We would subtract a bit of time from $\tfo$ because we know that only the twentieth molecule to reach $\ketS$ gives $t_{\rm c}$, whereas all hundred molecules in the sample contribute to $\tfo$, including the very slow ones that increase the average hitting time. We have also considered $t_{\rm c}$ for other values of $\eta$ and found that the step-like shape of Fig.~\ref{t41andtc}~(b) is retained and the $t_{\rm c}$ curves always lie below the $t_{41}$ curve. We show how the hitting times are distributed in the quantum and classical limits in Figs.~\ref{f41}~(a) and (b) respectively. Note that in Fig.~\ref{f41} we have plotted $f_{41}$ against $n$ [recall \eqref{f41Final}], the number of time steps taken to reach state $\ketS$ from $\ketg$. The hitting time is simply $n\,\dt$. 

As can be seen in Fig.~\ref{f41}~(a), the hitting-time distribution oscillates in the quantum limit. For clarity we have superimposed the quantum and classical limits of the distribution over a smaller range of $n$ containing the oscillations in Fig.~\ref{f41}~(c). We have stated the strength of the dephasing noise in the inset by quoting both $\mutt$ and its rate $q_{32}$ since we have quoted the strength of all other processes in the system by their rates. The average number of time steps taken to reach state $\ketS$ is also shown in the inset. One might wonder why $f_{41}(n)$ oscillates since it is actually a distribution of times rather than state-occupation probabilities (which is the quantity that one associates oscillatory motion to quantum behaviour). Nevertheless, $f_{41}(n)$ is in the end a transition probability, expressed by \eqref{f41Defn}, and we can understand why it oscillates by using the intuition gained from calculating the transition probability \eqref{PrQjQk}. Equation \eqref{PrQjQk} describes purely coherent evolution and oscillates indefinitely. The conditional probability defined in \eqref{f41Defn} is similar to \eqref{PrQjQk} apart from the extra conditioning required to make $f_{41}(n)$ a hitting-time distribution. Apart from how they are defined, the time evolution operators used to calculate these two transition probabilities are also different. In \eqref{f41Defn} we have used the map \eqref{Kb} which includes both coherent and decoherent parts. The coherent part is described by a unitary operator ($\hat{U}_{32}$) which generates oscillations in similar fashion as \eqref{PrQjQk}. The decoherent part in \eqref{Kb} (amplitude damping and dephasing) then acts to reduce the oscillations giving the net result seen in Fig.~\ref{f41}. By increasing $\mutt$ gradually we have found that the oscillations persist for $\mu_{32}$ values up to $10^{-7}$ and begin to die out for $\mu_{32} \approx 10^{-6}$ or above. This is consistent with the onset of the quantum-to-classical transition seen in Fig.~\ref{t41andtc}~(a). For times on the order of $10^{-8}\,{\rm s}$ (corresponding to the order of $t_{\rm c}$), $n$ is on the order of $10^6$. It can be seen in both Figs.~\ref{f41}~(a) and (b), that this is much less than the average hitting time, and is in fact even lower than the most probable time which is defined by the value of $n$ at which $f_{41}$ peaks [shown as $n_{\rm mp}$ in Figs.~\ref{f41}~(a) and (b)]. This is why $t_{\rm c}$ is much less than $t_{41}$ in Figs.~\ref{t41andtc}~(a) and (b).

The evolution of the system state is shown in Fig.~\ref{rho}. In Figs.~\ref{rho}~(a)--(c) we plot the occupation probabilities of $\ketg$, $\kets$, and $\ketS$ as functions of $n$ for different values of the dephasing noise (shown in the inset). For clarity we have omitted the triplet occupation probability $\rho_{33}$. As said in the first point under \eqref{Kb}, there are no other coherences in the system except for $\rho_{32}$. We show its magnitude as a function of $n$ for different $\mutt$ values in Fig.~\ref{rho}~(d). Comparing Figs.~\ref{rho}~(a)--(c) it is clear that the oscillations in $\rho_{11}$ and $\rho_{22}$ are strongest in Fig.~\ref{rho}~(a). These are essentially Rabi oscillations which are well-known in the study of atom-photon interactions except that our model does not refer explicitly to such a system. When we increase the dephasing by one order of magnitude the oscillations are visibly reduced [Fig.~\ref{rho}~(b)] and eventually vanish when $\mutt$ is further increased by another order of magnitude [Fig.~\ref{rho}~(c)]. Note the values associated with the disappearance of the oscillations in Figs.~\ref{rho}~(b) and (c) are consistent with the onset of the quantum-to-classical region in $\tfo$. From Fig.~\ref{t41andtc}~(a) we see that the quantum-to-classical transition starts somewhere around $10^{-6}$--$10^{-5}$ and this is also when coherent oscillations in $\rho_{22}$ start to disappear. Furthermore, the sensitivity of the average hitting time to $\mutt$ in the quantum-to-classical region can also be seen in the decay of the coherences in the system shown in Fig.~\ref{rho}~(d): The reduction in $|\rho_{32}|$ when going from $\mutt=10^{-6}$ to $\mutt=10^{-5}$ is much greater than the reduction when going from $\mutt=10^{-7}$ to $\mutt=10^{-6}$. If we accept that coherences in the system can speed up a quantum walk then the rise in $\tfo$ seen in the quantum-to-classical segment of Fig.~\ref{t41andtc} can be attributed to rate at which coherences are lost as shown in Fig.~\ref{rho}~(d). It is also interesting to note that while the occupation probabilities at $\mutt=10^{-5}$ [Fig.~\ref{rho}~(c)] do not exhibit oscillations at all, the system still has some coherence as shown by the black dashed curve in Fig.~\ref{rho}~(d). In the next paragraph we provide a sketch of the occupation probabilities shown in Figs.~\ref{rho}~(a)--(c).

Our model does not include states which trap a molecule indefinitely (apart from $\ketS$) or include losses so the total number of molecules is conserved. This means that a hundred percent of the molecules must eventually reach the triplet-product state. Hence the occupation probability of the triplet-product state must approach one in the long-time limit whereas all the other states must approach zero. Since the triplet-product state is initially unpopulated and each molecule can only make transitions towards it, we find that $\rho_{44}$ is a monotonically increasing function of time starting at zero. Similarly every molecule in the ensemble starts at state $\ketg$ and must eventually leave this state so one expects there to be an initial decay in $\rho_{11}$ starting at one. The rate of decay of $\rho_{11}$ will depend on the precise values of the transition rates, particularly the values of $k_{21}$ and $k_{12}$. For $k_{21}>k_{12}$ the decay in $\rho_{11}$ is steeper than if $k_{21}<k_{12}$ (not shown). We can see from Fig~\ref{rho}~(a) that $\rho_{11}$ oscillates even though $\ketg$ does not participate directly in any coherent transitions. The reason is because it is coupled to $\kets$, so that the time dependence of $\rho_{11}$ is affected by how $\rho_{22}$ depends on time \footnote{More precisely, the time dependence of $\rho_{11}$ depends on the time integral of $\rho_{22}$, which evaluates to an oscillatory function when $\rho_{22}$ is an oscillatory function.}. Of course one could then ask why $\rho_{44}$ does not oscillate in Fig.~\ref{rho} since it is also coupled to $\rho_{22}$. In principle this is possible but whether oscillations actually occur in $\rho_{44}$ will also depend on the precise values of the transitions rates. We have tested this by decreasing $k_{42}$ from $3.3 \times 10^6\,{\rm s}^{-1}$ to $3.3\times10^4\,{\rm s}^{-1}$ and observed that $\rho_{44}$ does indeed oscillate (not shown). However, we note that oscillations in $\rho_{44}$ are such that it always remains a monotonically increasing function. This is consistent with Fig.~\ref{ChemReactGraph} in which there are no transitions out of $\ketS$, only transitions into it. Similarly the oscillations seen in $\rho_{11}$ in Fig.~\ref{rho}~(a) are also consistent with the fact that we have allowed for transitions back to $\ketg$ from $\kets$. Finally, we mention that if oscillsations are absent in $\rho_{22}$ then it must always start from zero and rise to a certain point followed by an eventual decay to zero again. We can understand this by noting that we have set the transition rates out of $\kets$ ($k_{42}$ and $k_{12}$ in Table~\ref{RatesSummary}) to be less than the rate going into $\kets$ ($k_{21}$ in Table~\ref{RatesSummary}). This means that there is a chance for the population of $\kets$ to build up to some critical value. From this value it must then decay to zero since sooner or later a molecule will make a transition to the triplet-product state and stay there. Everytime a molecule reaches $\ketS$ a lesser amount of molecules is left behind to be distributed between the remaining states. This is why $\rho_{22}$ (and the occupation probability of the other states) must eventually decay to zero. We have assumed for simplicity that oscillations in $\rho_{22}$ are absent. However, if oscillations are present in $\rho_{22}$ then our description is one of its envelope as a function of time. A similar sort of reasoning can be applied to $\rho_{33}$ so we will not bother explaining it.

\subsection{Results for unrealistic rates}
\label{UnrealRates}

\begin{figure*}
\center
\includegraphics[width=0.5\textwidth]{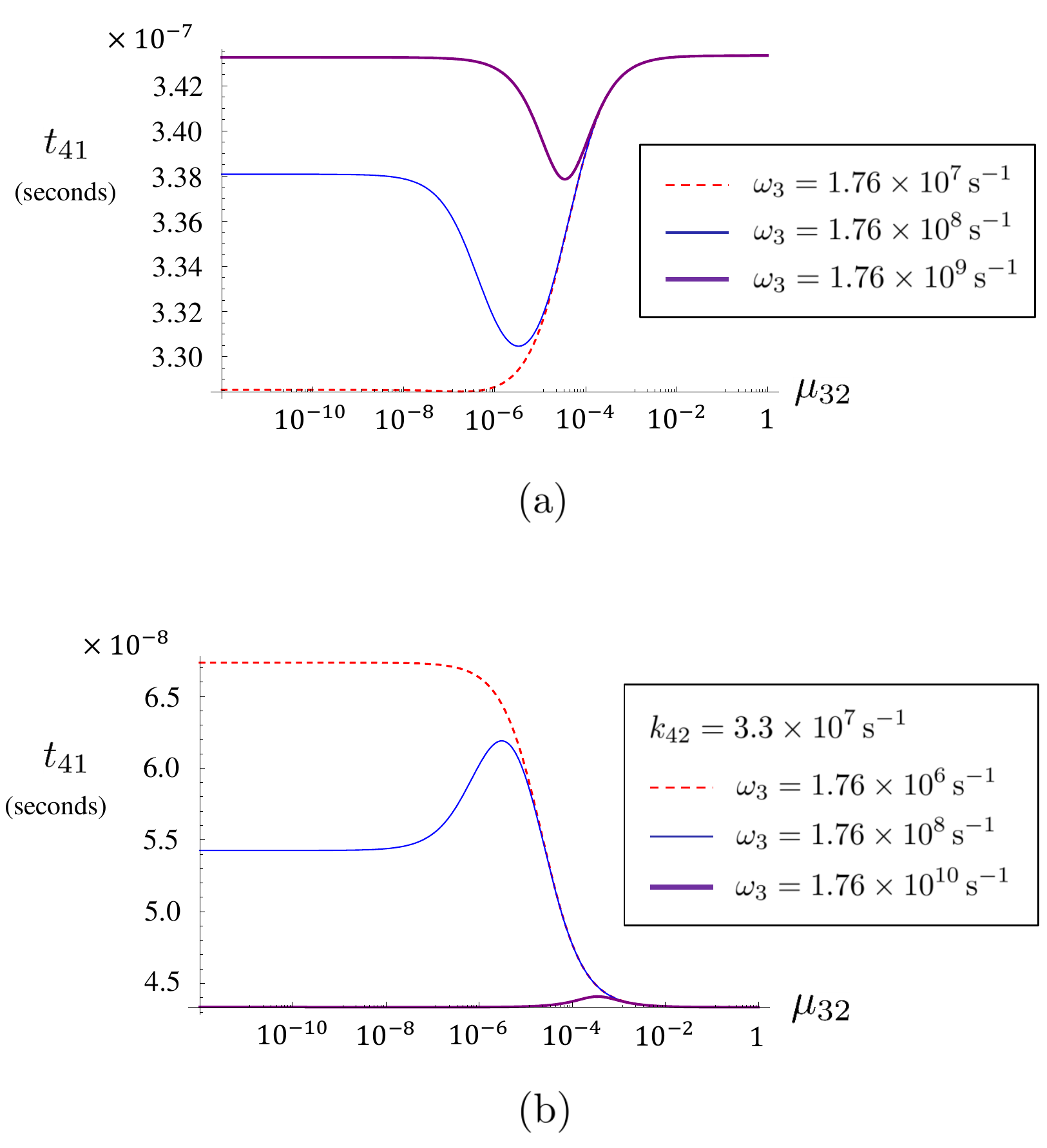} 
\caption{The average hitting time as a function of the dephasing noise in log scale when $\omega_3$ has the values shown in the inset. (a) When $\omega_3$ becomes greater than its nominal value corresponding to the dashed red curve, the molecule is biased towards taking the route via $\kets$ in Fig.~\ref{ChemReactGraph} as explained in the main text. (b) The value of $k_{42}$ is an order of magnitude larger than its value in Table~\ref{RatesSummary} and when $\omega_3$ takes on much larger values shown in the inset.}
\label{t41NonEstA}
\end{figure*}
We saw in Fig.~\ref{t41andtc} that for the system parameters in Table~\ref{RatesSummary} the average hitting time makes a step-like transition from the quantum to classical regime. Although we managed to estimate the order of magnitude of $\tfo$, its behaviour as a function of the dephasing noise remains nontrivial. The dependence of the average hitting time on dephasing will change if the rates in Table~\ref{RatesSummary} have different values. We illustrate this point in Fig.~\ref{t41NonEstA}. In Fig.~\ref{t41NonEstA}~(a) we change only $\omega_3$ by making it significantly larger than its value in Table~\ref{RatesSummary} (shown in the inset). For reference we reproduce Fig.~\ref{t41andtc}~(a) which is shown as the dashed red curve. It can then be seen from the thin blue curve and the thick purple curve that as we increase $\omega_3$ the average hitting time increases in the quantum regime while it remains constant in the classical regime. Most noticeably a dip is formed in the process. This shows the existence of a unique value of dephasing noise that minimizes the average hitting time and thereby showing that a more coherent system does not always lead to faster reaction. A similar result have been reported for the transport of excitons in photosynthesis in Refs.~\cite{PH08,RMKLAG09} from a microscopic description using master equations. Here we have arrived at the same conclusion by describing the state transitions phenomenologically using Kraus maps. The size of the dip decreases as $\omega_3$ is increased until eventually the average hitting time becomes flat when $\omega_3$ approaches infinity. The vanishing of the dip can be understood by first realising that increases in $\omega_3$ decreases the transition probability to state $\kett$ from $\kets$. This can be seen from \eqref{PrQjQk} where $\omega_3$ appears only in the denominator. Thus in the limit of $\omega_3 \longrightarrow \infty$, the system behaves as if $\kett$ is nonexistent and the only route the system can take to reach $\ketS$ is via $\kets$. The random walk then becomes independent of $\mutt$ and can be regarded as a classical three-state walk. We do not have a simple explanation for why the minimum occurs in Fig.~\ref{t41NonEstA}~(a).

In Fig.~\ref{t41NonEstA}~(b) we explore the case when the singlet state $\kets$ decays to the triplet-product state in a much shorter time than the triplet does. We thus set the rate $k_{42}$ to be one order of magnitude bigger than $k_{43}$ and considered the values of $\omega_3$ shown in the inset. Here we have the ``reverse'' situation of Fig.~\ref{t41NonEstA}~(a): The average hitting time is now longer in the quantum regime than it is in the classical (except for the large-$\omega_3$ limit). For intermediate values of $\omega_3$ the minimum seen in Fig.~\ref{t41NonEstA}~(a) has ``turned into'' a maximum. As with Fig.~\ref{t41NonEstA}~(a), the large-$\omega_3$ limit is equivalent to a three-state classical walk and independent of $\mutt$. Hence we find in Fig.~\ref{t41NonEstA}~(b) that $\tfo$ becomes flat and approaches the value in the classical limit. Just as we were unable to offer a simple explanation for the occurrence of the minimum seen in Fig.~\ref{t41NonEstA}, here we do not understand the appearance of the maximum in Fig.~\ref{t41NonEstA}~(b). However, we offer a plausible explanation as to why $\tfo$ attains a larger value in the quantum limit than in the classical limit. Suppose first that our random walk is fully quantum. Starting at state $\ketg$ the system will eventually make a transition to $\kets$. Once it reaches $\kets$, the system starts to have a probability amplitude that is spread over both states $\kets$ and $\kett$. This in turn means the decay to $\ketS$ from both $\kets$ and $\kett$ will contribute to $\tfo$. Now compare this to the case when the random walk is fully classical. In this case the probability amplitude is localised to one state at a time. Starting again at $\ketg$, the system makes a transition to $\kets$. Because we have set $k_{42}$ to be much greater than $k_{43}$, most of the time the system will jump to $\ketS$ from $\kets$. Only on a few occasions will the transition to $\ketS$ be from $\kett$. Thus the dominant contribution to $\tfo$ in the classical limit will come from the $\ketg \longleftrightarrow \kets$ transition, and the $\kets \longrightarrow \ketS$ transition. This makes the classical average hitting time shorter than the quantum one because the quantum calculation takes into account the time it takes to go through $\kett$, which takes a much longer time to reach $\ketS$ because $k_{43}$ is much less than $k_{42}$. This is also consistent with the ordering of $\tfo$ in the quantum limit for different values of $\omega_3$. That is, for $\mutt \longrightarrow 0$, the red dotted curve is above the thin blue curve because $\omega_3$ for the red dotted curve is much smaller than the $\omega_3$ for the thin blue curve. Having a smaller $\omega_3$ means a greater spread of the probability amplitude across $\kett$ and $\kets$. This leads to a greater contribution to $\tfo$ coming from the route via $\kett$, which takes more time. For the same reason we find the thin blue curve to be above the thick purple curve in Fig.~\ref{t41NonEstA}~(b).

\section{Conclusion}
\label{Discussion}

We have studied the radical-pair reaction from the viewpoint of coherent chemical kinetics where the transient populations in the reaction and the reaction time are obtained using an approach analogous to classical rate equations. This approach can be said to be one of quantum walks because the analogous classical model falls under the well-known theory of Markov chains which is essentially a theory of classical random walks. The quantum-walk approach was explained in detail in an earlier paper (Part I) so the objective here is to apply the quantum-walk idea to an example with realistic intermediate transition rates. Besides just constructing the time-evolution map for describing the reaction, we have also shown how other quantities such as the reaction time can be calculated using the quantum-walk approach.

For the reaction modelled by Fig.~\ref{ChemReactGraph}, with its transition rates given in Table~\ref{RatesSummary}, we found its reaction time to be essentially a classical property [Fig.~\ref{t41andtc}~(a)] but not necessarily its populations [Fig.~\ref{rho}~(a)]. We have attributed this result to the final steps taking $\ket{\psi_2}$ and $\ket{\psi_3}$ to $\ket{\psi_4}$ in the reaction. These transitions are rate limiting because they have a significantly slower transition rate than all other transitions and they also happen to be incoherent processes. The independence of the hitting time on the coherence of the quantum walk can also be seen in the hitting-time distribution where only mild oscillations are produced in the quantum limit (Fig.~\ref{f41}). Whether the reaction population dynamics can be effectively treated as classical will depend on the actual value of dephasing used in the model. When an accurate estimate of this is known the quantum-walk model can then be used to benchmark the quality of a classical rate-equations model where coherences are ignored. We emphasise again that our results on the dependence of the radical-pair kinetics on coherences are for a given magnetic field. Other studies on how quantum coherence (or decoherence) might play a role in the radical-pair model for its function as a compass (e.g.~its directional sensitivity) have been explored elsewhere \cite{HBH12,CCP12,TB12,CP13,PZBK13,GRMBV11}.

In light of the result obtained here, one possible avenue of future work is modelling the transtion from \RPone\ to \RPthr\ of Fig.~\ref{f1} in Appendix~\ref{CryReact}. Since we already know that this sequence of transitions occur extremely fast it would be interesting to consider the average hitting time of the $\RPone \longrightarrow \RPthr$ transition as a function of the coherence in each intermediate radical pair.

\section{Acknowledgement}

We would like to thank Vlatko Vedral and Ata\c{c} $\dot{\rm I}$mamo$\breve{\rm g}$lu for useful discussions. TP acknowledges support from the Start-Up grant of the Nanyang Technological University and Ministry of Education grant number RG127/14. AC, PK, and DK acknowledge support from the National Research Foundation and Ministry of Education in Singapore.

\section*{Appendices}
\appendix

\section{Radical pairs in \emph{Arabidopsis thaliana}}
\label{CryReact}

\subsection{A realistic reaction in {\em Arabidopsis thaliana} cryptochromes}

The simple scheme of Fig.~\ref{f3} can be seen to arise from the radical-pair formation model of cryptochromes in \emph{Arabidopsis thaliana} as described in Ref.~\cite{SCS07}. Cryptochromes are a class of photoreceptor signalling proteins whose magnetic-sensing ability was first suggested by Ritz and coworkers \cite{RAS00}. Evidence for cryptochromes as a viable magnetoreceptor has been reviewed extensively (see for example Refs.~\cite{LM09,WW14}).

\begin{figure*}[t]
\centerline{\includegraphics[width=0.7\textwidth]{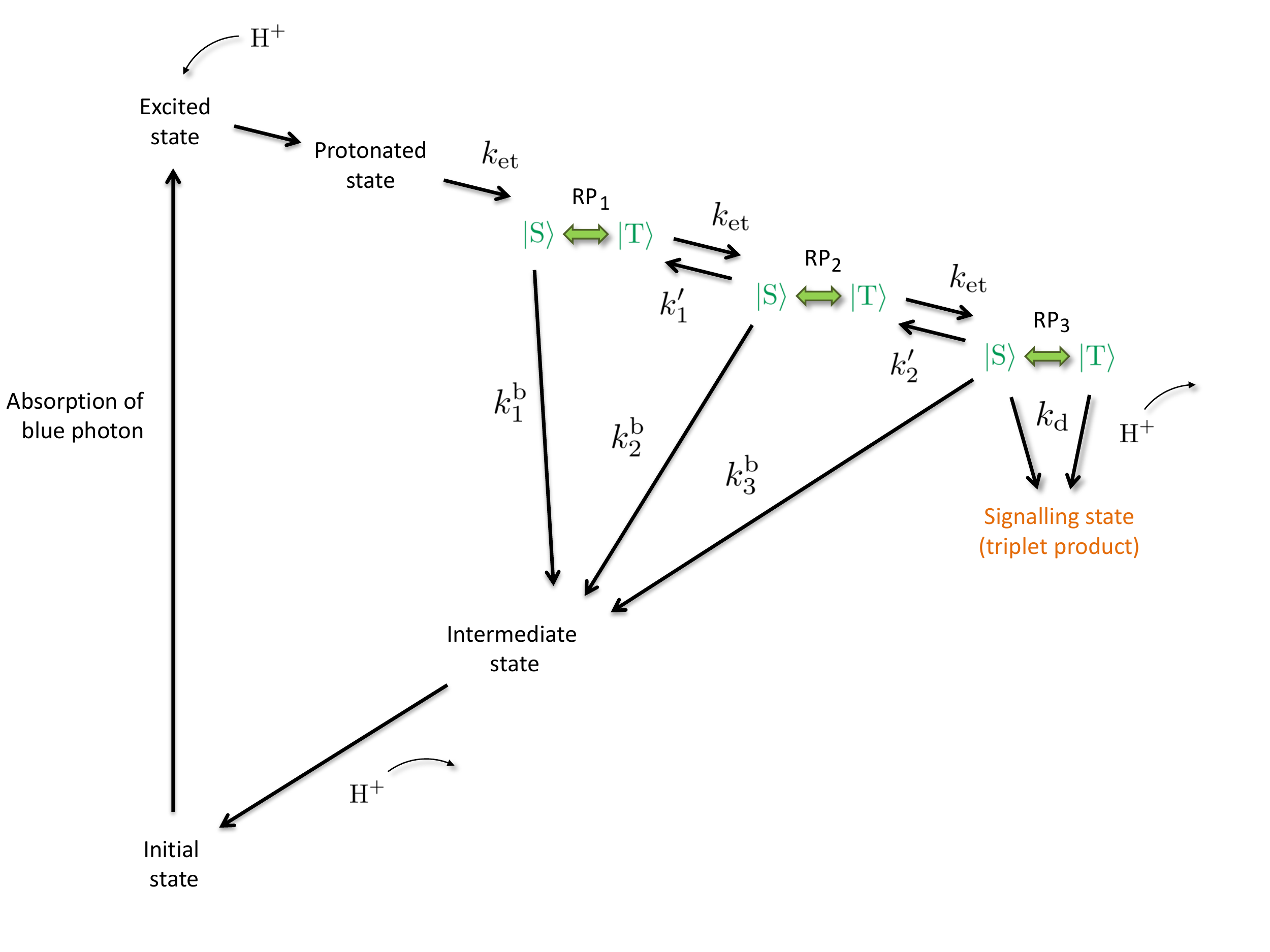}}
\caption{\label{f1} Schematic diagram of the radical-pair reaction path as described in Ref.~\cite{SCS07}. The initial state is excited by a blue photon and protonated. The protonation triggers a sequence of electron transfers in the molecule. The path consists of three radical pairs with \RPthr\ being the radical pair corresponding to Fig.~\ref{f3}. The electron transfer rates are taken from Ref.~\cite{SCS07}.}
\end{figure*}

The radical pair in Fig.~\ref{f3} is shown as \RPthr\ in Fig.~\ref{f1} and is formed as follows: The molecule in the initial state is excited by blue light and protonated. The protonation (shown as $\rm{H}^+$) triggers the electron transfer in the molecule that leads to a chain of radical pairs formed in sequence. The first radical pair (\RPone) is formed from the protonated state. From there it can evolve further forming a second radical pair (\RPtwo). If \RPone\ is in the singlet state it can return to the initial state. \RPtwo\ can evolve in the following ways: It can evolve forward forming a third radical pair (\RPthr) or evolve back to \RPone. If \RPtwo\ is in the singlet state the molecule can return to the initial state. Finally \RPthr\ can return either to \RPtwo, or if it is in the singlet state, to the initial state. The third way of evolution for \RPthr\ is deprotonation---the molecule forms the signalling state, which is the triplet product in Fig.~\ref{f3}. Here we will refer to this state as the triplet product or signallng state interchangeably. This can happen from both the singlet and triplet states of \RPthr. The amount of signalling state depends on the intensity and direction of the ambient magnetic field and carries the compass information. 

In Ref.~\cite{SCS07} the authors base their calculations on the known values of the different transfer rates. Many of the mentioned processes are neglected. First the lifetimes of \RPone\ and \RPtwo\ are shorter than the singlet-triplet interconversion time, therefore there is no significant change in the state of \RPone\ and \RPtwo. Significant singlet-triplet interconversion occurs only in the last radical pair (\RPthr), and the exchange and dipolar interactions can be neglected here due to the spatial separation between the radicals \cite{SDS13}. The rates for the transitions from \RPtwo\ to \RPone\ and \RPthr\ to \RPtwo\ are small so they are also neglected. In our calculations we use the simplified model of Fig.~\ref{f3}. Since \RPone\ and \RPtwo\ have very short lifetimes and do not have significant singlet-triplet interconversion we model the whole chain of reactions from the initial state to \RPthr\ as one process. The timescale of this process is taken to be around 10\,ns \cite{SCS07}. Note that other values can also be found in the literature which estimate this process to occur on the order of picoseconds or less, so the 10\,ns used by us is a very modest estimate \cite{DHW13}.

\subsection{Correspondence to rates used in the quantum-walk model}

\begin{table*}
	\begin{center}
		\begin{tabular}{c c}
		\hline
  	\hline
		Rate (Corresponding quantity in Ref.~\cite{SCS07} or Fig.~\ref{f1}) & Value used in simulation (${\rm s}^{-1}$) \\    
    \hline 
    $k_{21}$ ($k_{\rm et}$)          &   $1 \times 10^8$  \\ 
    $k_{12}$ ($k_{3}^{\rm b}$)       &   $1 \times 10^{7}$  \\ 
    $k_{43},\,k_{42}$ ($k_{\rm d}$)  &   $3.3 \times 10^6$  \\
		$\omega_3$ ($|E_3|$)             &   $1.76 \times 10^7$  \\
		$\omega_2$ ($E_1$)               &   $0$  \\
    $\Omega_{32}$ ($|V_{1\to3}|$)    &   $4.06 \times 10^7$  \\
		$1/\dt$                          &   $1 \times 10^{14}$  \\
		$q_{32}$                           &   $[0, 10^{14}]$  \\
  	\hline
  	\hline
		\end{tabular}
	\end{center}
	\caption{\label{RatesSummaryApp}Summary of rates used to simulate the graph of Fig.~\ref{ChemReactGraph}. The corresponding rates in Fig.~\ref{f1} and Ref.~\cite{SCS07} are shown in brackets. Note that $E_3$, $E_1$, and $V_{1\to3}$ are not shown in Fig.~\ref{f1} but can be found in Ref.~\cite{SCS07}.}
\end{table*}
The correspondence between the rates used in Fig.~\ref{ChemReactGraph} and Fig.~\ref{f1} are summarised in Table~\ref{RatesSummaryApp} and are as follows. We take the rate of the $\ketg \longrightarrow \kets$ transition in Fig.~\ref{ChemReactGraph} to be approximated by $k_{\rm et}$ in Fig.~\ref{f1}. This was measured in Refs.~\cite{BSE+04} and \cite{AVMEB00} to be $10^{8}\,{\rm s}^{-1}$ but for photolyase \cite{SCS07}. Due to the lack of data for cryptochrome we will use this value in our simulations as was done in Ref.~\cite{SCS07}. Therefore we set $k_{21}=10^8\,{\rm s}^{-1}$. Note that if $10^8\,{\rm s}^{-1}$ is an accurate estimate of $k_{\rm et}$ for cryptochrome then $k_{21}$ should be smaller than this number since we have neglected the initial stages of photon absorption, protonation, and the intermediate radical pairs. Thus $10^{8}\,{\rm s}^{-1}$ is likely to be an upper bound for the true value of $k_{21}$ in Fig.~\ref{ChemReactGraph}. As this number is for photolyase rather than cryptochrome we also consider the average hitting time when $k_{21}$ is an order of magnitude above and below $10^8\,{\rm s}^{-1}$ \cite{SCS07}. The $\kets \longrightarrow \ketg$ transition rate can be taken to be the rate at which the singlet state of \RPthr\ in Fig.~\ref{f1} decays to the initial state. We take this to be roughly the same order of magnitude as the rate $k^{\rm b}_3$ in Ref.~\cite{SCS07} which was estimated to be $10^7 \,{\rm s}^{-1}$. We thus set $k_{12}=10^7 \,{\rm s}^{-1}$. Since \RPthr\ makes transitions to the signalling state via deprotonation, we set the rates for the $\kets \longrightarrow \ketS$ and $\kett \longrightarrow \ketS$ transitions to be the same as the deprotonation rate $k_{\rm d}$ in Fig.~\ref{f1}. This gives $k_{42}=k_{43}=3.3 \times 10^6\,{\rm s}^{-1}$. Next we have the $\kets \longleftrightarrow \kett$ transition which corresponds to the singlet-triplet interconversion. Recall from Sec.~\ref{CohEvoDeph} that we have approximated the rate of this process in terms of the matrix elements of the Hamiltonian \eqref{Hhatjk}, given by $2 \zeta_{32}$ [see \eqref{ParameterDefnB}]. Estimates of the matrix elements of the Hamiltonian \eqref{Hhatjk} for $j=3$ and $k=2$ for an Earth-strength magnetic field can be found in Ref.~\cite{SCS07}. The values of $\omega_3$ and $\omega_2$ were estimated to be $-1.76 \times 10^7 \, {\rm s}^{-1}$ (quoted as $-1.158 \times 10^{-8}\,{\rm eV}$ in Ref.~\cite{SCS07}) and $0$ respectively. The difference $\omega_3-\omega_2$ then corresponds to the Zeeman splitting of the singlet and triplet states. The value of $\Omega_{32}$ is determined from physical constants related to the magnetic interactions and was estimated to be $-4.06 \times 10^7 \,{\rm s}^{-1}$ (quoted as $-2.674 \times 10^{-8} \,{\rm eV}$ in Ref.~\cite{SCS07}). This gives a singlet-triplet interconversion rate of $2\,\zeta_{32}=8.3 \times 10^7 \,{\rm s}^{-1}$. Note that having $\omega_2=0$ and $\omega_3$, $\Omega_{32}$ negative just changes the sign of the exponent in $\Uhat_{32} = \exp(-i \Hhat_{32} \dt)$ so for simplicity we will take the matrix elements of $\hat{H}_{32}$ to be the absolute values of the above numbers. Lastly, an accurate simulation of the time evolution defined by \eqref{rho(tn)} requires a time step $\dt$ that is much smaller than any of the rates above. A method to find a suitably small value of $\dt$ is to use the independence of the average hitting time $\tfo$ on $\dt$ for a fixed set of system parameters. For the parameters shown in Table~\ref{RatesSummaryApp} we find that $\tfo$ stops changing when $\dt$ is $10^{-13}\,{\rm s}$ or less. We therefore set $\dt=10^{-14}$ which means that the dephasing parameter $\mutt$ (which varies between zero and one) can be specified by a rate $q_{32}$ which varies between $0$ and $10^{14}$ as defined in \eqref{DecParam}. We will use \eqref{DecParam} to make the rate of dephasing explicit for plots that use a fixed value of $\mutt$ since we have specified the strength of all other processes by specifying its rate of occurrence. We note also that an estimate of the singlet-triplet dephasing rate using a first-principles calculation was recently published in Ref.~\cite{Wal14}.

\section{Exact form of $\hat{U}_{jk}$}
\label{AppA}

To derive the explicit form of the unitary operator $\Uhat_{jk}(t)$ in the basis $\{\ket{\psi_k}\}_{k=1}^N$ for any $t$ we first recall that the Hamiltonian is given by 
\begin{align}
\label{HhatjkAppendix}
	\Hhat_{jk} = \omega_j \, \Qj + \omega_k \, \Qk + \Omega_{jk} \big( \Qjk + \Qkj \big)  \;,
\end{align}
where $\Qj = \op{\psi_j}{\psi_j}$ and $\Qjk = \op{\psi_j}{\psi_k}$. It is clear that for any $j \ne k$,
\begin{gather}
\label{ProjP1}
	\Qjk \Qkj = \Qj \;, \\ 
\label{ProjP2}
	\Qjk \Qk = \Qjk   \;, \\ 
\label{ProjP3}
	\Qjk \Qjk = \Qjk \Qj = \Qk \Qj = 0  \;. 
\end{gather}
It will be convenient to write the Hamiltonian in terms of the sum and difference frequencies $\sigma_{jk}$ and $\Delta_{jk}$ defined as
\begin{align}
\label{ParameterDefnA}
	\sigma_{jk} = \frac{1}{2} \big( \omega_k + \omega_j \big)  \;, \quad  	\Delta_{jk} = \frac{1}{2} \big( \omega_k - \omega_j \big)
\end{align}
The first two terms in \eqref{HhatjkAppendix} can then be written as
\begin{align}
	\omega_j \, \Qj + \omega_k \, \Qk = {}& (\sigma_{jk} - \Delta_{jk}) \, \Qj + (\sigma_{jk} + \Delta_{jk}) \Qk  \nn \\
                                    = {}& \sigma_{jk} \, (\Qk + \Qj)  + \Delta_{jk} \, (\Qk - \Qj)  \;.
\end{align}
The unitary operator is thus
\begin{align}
\label{Uappendix1}
	\Uhat_{jk}(t) = e^{-i\Hhat_{jk}t} 
	              = {}& \exp\!\big\{ \!-i \big[ \, \sigma_{jk} \, (\Qk + \Qj)  \nn \\
	                  & + \Delta_{jk} \, (\Qk - \Qj)  \nn \\
	                  & + \Omega_{jk} ( \Qjk + \Qkj ) \big] \, t \, \big\}  \;.
\end{align}
Note that 
\begin{align}
	\big[ \Qj+\Qk, \Qj-\Qk \big] = {}& \big[ \Qj+\Qk, \Qjk+\Qkj \big] \nn \\
	                             = {}& 0  \;,
\end{align}
so that \eqref{Uappendix1} can be factored as
\begin{align}
\label{Ufactorised}
	\Uhat_{jk}(t) = {}& \exp\!\big\{ \!-\!i \big[ \sigma_{jk} (\Qj + \Qk) \big] t \big\}  \nn \\
	                  & \times \exp\!\big\{ \!-\!i \big[ \Delta_{jk} \, (\Qk - \Qj)  \nn \\ 
	                  & + \Omega_{jk} ( \Qjk + \Qkj ) \big] \, t \, \big\}  \;.
\end{align}

The first factor can be simplified by noting that $\Qj+\Qk$ is the projector onto the subspace spanned by $\ket{\psi_j}$ and $\ket{\psi_k}$ so that we have, for any integer $n>0$,
\begin{align}
	(\Qj + \Qk)^n = \Qj + \Qk  \;.
\end{align}
This gives
\begin{align}
	{}& \exp\!\big[\!-\!i \sigma_{jk} (\Qj + \Qk) \, t \, \big]   \nn \\
	  & = \hat{1} + \sum_{n=1}^{\infty} \, \big(\Qj + \Qk\big)^n \frac{(-i\sigma_{jk}\,t)^n}{n!}  \nn \\
    & = \hat{1} + \big(\Qj + \Qk\big) \sum_{n=1}^{\infty} \, \frac{(-i\sigma_{jk}\,t)^n}{n!}  \nn \\
    & = \hat{1} + \big(\Qj + \Qk\big) \big( e^{-i\sigma_{jk}\,t} - 1 \big)  \nn \\
\label{Ufac1}
   & = \Pjk + \big(\Qj + \Qk\big) \, e^{-i\sigma_{jk}\,t}  \;,
\end{align}
where we have defined 
\begin{align}
	\Pjk = \hat{1} - \big( \Qj + \Qk \big)  \;.
\end{align}
This projects the system to states which are not spanned by $\ket{\psi_j}$ and $\ket{\psi_k}$ and thus satisfies  
\begin{gather}
\label{ProjP4}
	  \Pjk^2 = \Pjk  \;, \\
\label{ProjP5}
	\Pjk \Qj = \Pjk \Qk = \Pjk \Qkj = \Pjk \Qjk = 0  \;.
\end{gather}

The second factor in \eqref{Uappendix1} can be simplified by a similar approach except here we require the $n$th power of $\Delta_{jk}(\Qk-\Qj) + \Omega_{jk}(\Qjk+\Qkj)$ where $n$ is any positive integer. This is given by
\begin{align}
	{}& \big[ \Delta_{jk} \big( \Qk - \Qj \big) + \Omega_{jk} \big( \Qjk + \Qkj \big) \big]^n  \nn \\
	& = \big( \Delta^2_{jk} + \Omega^2_{jk} \big)^{[n-f(n)]/2}  
	    \Big\{ \big[ 1 - f(n) \big] \big( \Qk + \Qj \big)^{1-f(n)}   \nn \\
	& \quad + f(n) \big[ \Delta_{jk} \big( \Qk - \Qj \big)  
	    + \Omega_{jk} \big( \Qjk + \Qkj \big) \big]^{f(n)}  \Big\}  \;,
\label{nthPower}
\end{align}
where we have defined the parity function
\begin{align}
\label{ParityFunc}
	f(n) = \frac{1+(-1)^{n+1}}{2} = \left\{ \begin{array}{c}  1 \,, \; n=1,3,5,\ldots.  \\ 
	                                                          0 \,, \; n=2,4,6,\ldots.  \end{array} \right.
\end{align}
We will prove \eqref{nthPower} in Appendix~\ref{AppB}. Note that because of $f(n)$ the Taylor series for the second exponential in \eqref{Ufactorised} will separate into a sum with only odd powers and a sum with only even powers. We thus have 
\begin{align}
	{}& \exp\!\big\{ \!-\!i \big[ \Delta_{jk} \big( \Qk - \Qj \big) + \Omega_{jk} \big( \Qjk + \Qkj \big) \big] t \big\}  \nn \\
	{}& = \hat{1} + \sum_{n=1}^\infty \, \frac{(-i t)^n}{n!} \big[ \Delta_{jk} \big( \Qk - \Qj \big)  
	      + \Omega_{jk} \big( \Qjk + \Qkj \big) \big]^n   \nn \\[0.25cm]
	{}& = \hat{1} + \sum_{n=0}^\infty \; (-i)^{2n+1} \, \frac{t^{2n+1}}{(2n+1)!} \;\big[ \Delta_{jk} \big( \Qk - \Qj \big)  \nn \\
	      & \quad + \Omega_{jk} \big( \Qjk + \Qkj \big) \big]^{2n+1}   \nn \\
	      & \quad + \sum_{n=1}^\infty \; (-i)^{2n} \, \frac{t^{2n}}{(2n)!} \; \big[ \Delta_{jk} \big( \Qk - \Qj \big)   \nn \\
	      & \quad + \Omega_{jk} \big( \Qjk + \Qkj \big) \big]^{2n}  \nn \\[0.25cm]
\label{Exp2}      
	{}& = \hat{1} - i \big[ \Delta_{jk} \big( \Qk - \Qj \big) + \Omega_{jk} \big( \Qjk + \Qkj \big) \big]  \nn \\
	      & \quad \times \sum_{n=0}^\infty \; (-1)^n \big( \Delta^2_{jk} + \Omega^2_{jk} \big)^n \frac{t^{2n+1}}{(2n+1)!} \nn \\
	    {}& \quad + \big( \Qk + \Qj \big) \sum_{n=1}^{\infty}\; (-1)^n \big( \Delta^2_{jk} + \Omega^2_{jk} \big)^n \frac{t^{2n}}{(2n)!}  \;.
\end{align}
The last equality follows from setting $n$ in \eqref{nthPower} to be $2n+1$ and $2n$ [see also \eqref{OddPower} and \eqref{EvenPower} in Appendix~\ref{AppB}]. It will be convenient to define
\begin{align}
\label{ZetajkAppendix}
	\zeta^2_{jk} = \Delta^2_{jk} + \Omega^2_{jk}  \;,
\end{align}
with $\zeta_{jk}$ taken to be the positive square root of $\Delta^2_{jk} + \Omega^2_{jk}$. We can then write \eqref{Exp2} as
\begin{align}
	{}& \exp\!\big\{ \!-\!i \big[ \Delta_{jk} \big( \Qk - \Qj \big) + \Omega_{jk} \big( \Qjk + \Qkj \big) \big] t \big\}  \nn \\
	{}& = \hat{1} - i \big[ \Delta_{jk} \big( \Qk - \Qj \big) + \Omega_{jk} \big( \Qjk + \Qkj \big) \big] 
	      \frac{1}{\zeta_{jk}} \nn \\
	      & \quad \times \sum_{n=0}^\infty \; (-1)^n \; \frac{\big( \zeta_{jk} \, t \big)^{2n+1}}{(2n+1)!} \nn \\
	      & \quad + \big( \Qk + \Qj \big) \sum_{n=1}^{\infty}\; (-1)^n \; \frac{\big( \zeta_{jk} \, t \big)^{2n}}{(2n)!}  \nn \\[0.25cm]
	{}& = \hat{1} + \big[ \cos\big( \zeta_{jk} t \big) - 1 \big] \big( \Qk + \Qj \big)  \nn \\
	      & \quad - i \; \frac{\sin\big( \zeta_{jk} t \big)}{\zeta_{jk}} \; \big[ \Delta_{jk} \big( \Qk - \Qj \big) 
	      + \Omega_{jk} \big( \Qjk + \Qkj \big) \big]  \nn \\[0.25cm]
\label{Ufac2}    
	{}& = \Pjk + \frac{1}{2} \; \Big( e^{i \zeta_{jk}t} + e^{-i\zeta_{jk}t} \Big) \big( \Qk + \Qj \big)  \nn \\
	      {}& \quad - \; \frac{1}{2\,\zeta_{jk}} \; \Big( e^{i \zeta_{jk}t} - e^{-i\zeta_{jk}t} \Big) \; \big[ \Delta_{jk} \big( \Qk - \Qj \big)  \nn \\
	        & \quad + \Omega_{jk} \big( \Qjk + \Qkj \big) \big] \;.
\end{align}
Substituting \eqref{Ufac1} and \eqref{Ufac2} into \eqref{Ufactorised} and using the projector properties \eqref{ProjP1}--\eqref{ProjP3}, \eqref{ProjP4}, and \eqref{ProjP5} we get
\begin{align}
	\Uhat_{jk}(t) = {}& \Big\{ \Pjk + \big(\Qj + \Qk\big) \, e^{-i\sigma_{jk}\,t} \Big\}  \nn \\
	                  & \times \Big\{ \Pjk + \frac{1}{2} \; \Big( e^{i \zeta_{jk}t} + e^{-i\zeta_{jk}t} \Big) \big( \Qk + \Qj \big)  \nn \\
	                  & - \frac{1}{2\,\zeta_{jk}} \; \Big( e^{i \zeta_{jk}t} - e^{-i\zeta_{jk}t} \Big) 
	                    \big[ \Delta_{jk} \big( \Qk - \Qj \big) \nn \\
	                  & + \Omega_{jk} \big( \Qjk + \Qkj \big) \big] \Big\}  \nn \\[0.25cm]
	          = {}& \Pjk + \frac{1}{2} \; \Big[ e^{-i(\sigma_{jk} - \zeta_{jk}) t} + e^{-i(\sigma_{jk} + \zeta_{jk}) t} \Big]  \nn \\ 
	              & \times \big( \Qk + \Qj \big)  
	                - \frac{\Delta_{jk}}{2\,\zeta_{jk}} \; \Big[ e^{-i (\sigma_{jk} - \zeta_{jk}) t} \nn \\
	              & - e^{-i (\sigma_{jk} + \zeta_{jk}) t} \Big] \big( \Qk - \Qj \big)  \nn \\
	              & - \frac{\Omega_{jk}}{2\,\zeta_{jk}} \; \Big[ e^{-i(\sigma_{jk}- \zeta_{jk})t}  \nn \\
	              & - e^{-i(\sigma_{jk} + \zeta_{jk}) t} \Big] \big( \Qjk + \Qkj \big)  \\[0.25cm]
            = {}& \Pjk + \frac{1}{2} \; \Bigg[ \bigg( 1 + \frac{\Delta_{jk}}{\zeta_{jk}} \bigg) e^{-i(\sigma_{jk}-\zeta_{jk})t}  \nn \\
                & +  \bigg( 1 - \frac{\Delta_{jk}}{\zeta_{jk}} \bigg) e^{-i(\sigma_{jk}+\zeta_{jk})t} \Bigg] \; \Qj  \nn \\
                & + \frac{1}{2} \; \Bigg[ \bigg( 1 - \frac{\Delta_{jk}}{\zeta_{jk}} \bigg) e^{-i(\sigma_{jk}-\zeta_{jk})t}  \nn \\
                & +  \bigg( 1 + \frac{\Delta_{jk}}{\zeta_{jk}} \bigg) e^{-i(\sigma_{jk}+\zeta_{jk})t} \Bigg] \; \Qk  \nn \\
                & - \frac{\Omega_{jk}}{2\,\zeta_{jk}} \; \Big[ e^{-i(\sigma_{jk}- \zeta_{jk})t} - e^{-i(\sigma_{jk} + \zeta_{jk}) t} \Big]  \nn \\ 
                & \times \big( \Qjk + \Qkj)  \;.
\end{align}

\section{Proof of (\ref{nthPower})}
\label{AppB}

The identity \eqref{nthPower} can be proven most easily by considering the odd and even powers separately. For odd powers we have
\begin{align}
\label{OddPower}
	{}& \big[ \Delta_{jk} \big( \Qk - \Qj \big) + \Omega_{jk} \big( \Qjk + \Qkj \big) \big]^{2m+1} \nn \\
	{}& = \big( \Delta^2_{jk} + \Omega^2_{jk} \big)^m \big[ \Delta_{jk} \big( \Qk - \Qj \big) + \Omega_{jk} \big( \Qjk + \Qkj \big) \big]  \;,
\end{align}
where $m=0,1,2,\ldots$. This can be proven by induction as follows
\begin{align}
	{}& \big[ \Delta_{jk} \big( \Qk - \Qj \big) + \Omega_{jk} \big( \Qjk + \Qkj \big) \big]^{2(m+1)+1} \nn \\
	{}& = \big[ \Delta_{jk} \big( \Qk - \Qj \big) + \Omega_{jk} \big( \Qjk + \Qkj \big) \big]^{2m+1}  \nn \\
	      & \quad \:\!\; \times \big[ \Delta_{jk} \big( \Qk - \Qj \big) + \Omega_{jk} \big( \Qjk + \Qkj \big) \big]^2   \nn \\[0.25cm]
	{}& = \big( \Delta^2_{jk} + \Omega^2_{jk} \big)^m \big[ \Delta_{jk} \big( \Qk - \Qj \big) + \Omega_{jk} \big( \Qjk + \Qkj \big) \big]  \nn \\
	      & \quad \:\!\; \times  \big( \Delta^2_{jk} + \Omega^2_{jk} \big) \big( \Qk + \Qj \big)  \nn \\[0.25cm]
	{}& = \big( \Delta^2_{jk} + \Omega^2_{jk} \big)^{m+1} \big[ \Delta_{jk} \big( \Qk - \Qj \big) + \Omega_{jk} \big( \Qjk + \Qkj \big) \big] \,,
\end{align}
where we have used the projector properties \eqref{ProjP1}--\eqref{ProjP3}. For even powers we have
\begin{align} 
\label{EvenPower}                                                                                      
	{}& \big[ \Delta_{jk} \big( \Qk - \Qj \big) + \Omega_{jk} \big( \Qjk + \Qkj \big) \big]^{2m}  \nn \\
	  & = \big( \Delta^2_{jk} + \Omega^2_{jk} \big)^m  \big( \Qk + \Qj \big) \;, 
\end{align}
where $m=1,2,3,\ldots$. This again can be shown by induction:
\begin{align}
	{}&\big[ \Delta_{jk} \big( \Qk - \Qj \big) + \Omega_{jk} \big( \Qjk + \Qkj \big) \big]^{2(m+1)}  \nn \\
	{}& = \big[ \Delta_{jk} \big( \Qk - \Qj \big) + \Omega_{jk} \big( \Qjk + \Qkj \big) \big]^{2m}  \nn \\
	  & \quad \:\!\; \times \big[ \Delta_{jk} \big( \Qk - \Qj \big) + \Omega_{jk} \big( \Qjk + \Qkj \big) \big]^2  \nn \\[0.25cm]
	{}& = \big( \Delta^2_{jk} + \Omega^2_{jk} \big)^m \big( \Qk + \Qj \big) \big( \Delta^2_{jk} + \Omega^2_{jk} \big) \big( \Qk + \Qj \big)  \nn	\\[0.25cm]
	{}& = \big( \Delta^2_{jk} + \Omega^2_{jk} \big)^{m+1} \big( \Qk + \Qj \big)  \;.
\end{align}
Combining \eqref{OddPower} and \eqref{EvenPower} into a single equation by using the parity function introduced in \eqref{ParityFunc} results in \eqref{nthPower}.

\section{Average hitting time}
\label{DerivationOfn41}

The procedure for calculating the hitting time is as outlined in Sec.~\ref{HittingTime}. The probability generating function $F_{41}(z)$ of $f_{41}(n)$ is  
\begin{align}
\label{F41Sum}
	{}&	F_{41}(z)  \nn \\
	  & = \sum_{n=0}^\infty \, f_{41}(n) \, z^n   \nn \\
    & = \sum_{n=1}^\infty \, {\rm Tr}\Big\{ \Qfour \, \mapK(\dt) \, \big[ \Pfour \, \mapK(\dt) \big]^{n-1} \rho(0) \Big\} \, z^n   \nn \\
    & = {\rm Tr}\left\{ \Qfour \, \mapK(\dt) \, z \, \left( \, \sum_{n=1}^\infty \, \big[ \Pfour \, \mapK(\dt) \big]^{n-1}  z^{n-1} \right) \rho(0) \right\}  \;.
\end{align}
The sum in the last line is a geometric series in $z \, \Pfour \, \mapK(\dt)$ and will converge to $\big[ \mathbbm{1} - z \, \Pfour \, \mapK(\dt) \big]^{-1}$ provided that $\mathbbm{1} - z \, \Pfour \, \mapK(\dt)$ is invertible. This will be the case if $\|z\,\Pfour\,\mapK(\dt)\| < 1$ where $\|{\cal A}\|$ denotes any norm of ${\cal A}$ \cite{HJ85}. Note that for any state $\rho$ represented by a $m \times m$ matrix, the superoperator equation ${\cal A} \rho$ will have a matrix representation where $\rho$ is represented by a $m^2 \times 1$ vector and ${\cal A}$ a $m^2 \times m^2$ matrix. This follows from a procedure in linear algebra known as vectorisation. The norm $\|{\cal A}\|$ can then be defined as a matrix norm using any of the existing definitions \cite{HJ85}. The generating function $F_{41}(z)$ therefore has a radius of convergence given by $|z| < 1/\,\|\Pfour\,\mapK(\dt)\|$. For such values of $z$ we thus have
\begin{align}
\label{F41Conv}
	F_{41}(z) = {\rm Tr}\Big\{ \Qfour \, \mapK(\dt) \, z \big[ \mathbbm{1} - z \, \Pfour \, \mapK(\dt) \big]^{-1} \, \rho(0) \Big\}  \;.
\end{align}
It can be seen from the normalisation of $f_{41}(n)$ that $\|\Pfour\,\mapK(\dt)\|<1$ so \eqref{F41Conv} is valid for any $z$ such that $|z|$ is between zero and some number greater than one. Physically this means that the molecule will hit $\ketS$ in a finite amount of time. Infinite hitting times are possible if, for example, there are absorbing states---states for which the molecule will stay in forever once they are reached.

Taking the derivative of \eqref{F41Conv} with respect to $z$ gives
\begin{align}
	{}& \frac{d}{dz} \, F_{41}(z)  \nn \\
	  & = {\rm Tr}\!\left\{ \Qfour \, \mapK(\dt) \, \left( \frac{d}{dz} \; z \big[ \mathbbm{1} - z \, \Pfour \, \mapK(\dt) \big]^{-1} \right) \, \rho(0) \right\}   \nn \\[0.25cm]
\label{dFdz}
    & = {\rm Tr} \bigg\{ \Qfour \, \mapK(\dt) \, \bigg( \big[ \mathbbm{1} - z \, \Pfour \, \mapK(\dt) \big]^{-1}   \nn \\
        & \quad \:\!\;  + \, z \, \Pfour \, \mapK(\dt) \big[ \mathbbm{1} - z \, \Pfour \, \mapK(\dt) \big]^{-2} \bigg) \, \rho(0) \bigg\}  \;.
\end{align}
There is no ambiguity in writing $\big[ \mathbbm{1} - z \, \Pfour \, \mapK(\dt) \big]$ to the power of $-2$ when the inverse of $\big[ \mathbbm{1} - z \, \Pfour \, \mapK(\dt) \big]$ exists (which is the case here). In this case powers of $-2$ can be taken to be either the square of the inverse or the inverse of the square. Setting $z=1$ in \eqref{dFdz},
\begin{align}
	n_{41} = {}& \left[ \frac{d}{dz} \, F_{41}(z) \right|_{z=1}  \nn \\[0.25cm]
         = {}& {\rm Tr} \bigg\{ \Qfour \, \mapK(\dt) \, \bigg( \big[ \mathbbm{1} - \Pfour \, \mapK(\dt) \big]^{-1}  \nn \\
             & + \Pfour \, \mapK(\dt) \big[ \mathbbm{1} - \Pfour \, \mapK(\dt) \big]^{-2} \bigg) \, \rho(0) \bigg\}  \nn \\[0.25cm]
	       = {}& {\rm Tr} \bigg\{ \Qfour \, \mapK(\dt) \, \bigg( \big[ \mathbbm{1} - \Pfour \, \mapK(\dt) \big]\big[ \mathbbm{1} - \Pfour \, \mapK(\dt) \big]^{-2}  \nn \\
             & + \Pfour \, \mapK(\dt) \big[ \mathbbm{1} - \Pfour \, \mapK(\dt) \big]^{-2} \bigg) \, \rho(0) \bigg\}  \nn \\[0.25cm]
         = {}& {\rm Tr}\!\left\{ \Qfour \, \mapK(\dt) \, \big[ \mathbbm{1} - \Pfour \, \mapK(\dt) \big]^{-2} \, \rho(0) \right\}  \;.
\end{align}
This is the average number of steps required to arrive at $\ketS$ for the first time. The actual average hitting time is simply
\begin{align}
	\tfo = n_{41} \, \dt  \;.
\end{align}

\end{document}